\documentclass[traditabstract]{aa} 

\usepackage{graphicx}
\usepackage{txfonts}

\begin{document}

\title{Stellar diffusion in barred spiral galaxies}

\author{Maura Brunetti\inst{1} \and Cristina
  Chiappini\inst{2} \and Daniel Pfenniger\inst{1}}

\institute{
Geneva Observatory, University of Geneva, CH-1290 Sauverny, 
Switzerland 
\and 
Leibniz-Institut f\"ur Astrophysik Potsdam (AIP), 
An der Sternwarte 16 D - 14482, Potsdam, Germany}

\authorrunning{M. Brunetti et al.}

\date{}

\abstract{We characterize empirically the radial
  diffusion of stars in the plane of a typical barred disk galaxy by
  calculating the local spatial diffusion coefficient
  and diffusion time-scale for bulge-disk-halo $N$-body
  self-consistent systems which initially differ in the 
  Safronov-Toomre-$Q_T$ parameter. We find different diffusion
  scenarios that depend on the bar strength and on the degree of
  instability of the disk. Marginally stable disks, with $Q_T \sim 1$,
  have two families of bar orbits with different values of angular
  momentum and energy, which determine a large diffusion in the
  corotation region. In hot disks, $Q_T> 1$, stellar diffusion is
  reduced with respect to the case of marginally stable disks. In cold 
  models, we find that spatial diffusion is not constant in time and
  strongly depends on the activity of the bar, which can move stars
  all over the disk recurrently. We conclude that to realistically
  study the impact of radial migration on the chemical evolution
  modeling of the Milky Way the role of the bar has to be taken into
  account.}
  

\maketitle

\section{Introduction}\label{introduction}

Disk galaxies are highly nonlinear systems which are driven by
external forcing (satellites, interaction with nearby galaxies),
internal instabilities (related to the formation of internal
structures such as spiral arms, rings, bars) or a combination of
both. Chaos and complexity, which are two different aspects of
nonlinear response, dominate the dynamics of galactic systems.  Disk
galaxies are `complex' in the sense that they are made of many
components (stars, gas, dark matter in the bar-bulge, disk and halo
components) whose interactions can give rise to spontaneous
self-organization and to the emergence of coherent, collective
phenomena. Examples of emergent behavior in disk galaxies are the
formation of a central bar or the onset of spiral arms and warps,
which can develop in dynamically cold disks (Revaz \& Pfenniger~\cite{revaz}).
Disk galaxies are not only complex systems, but even display chaotic
behavior, since tiny differences in initial orbits of stars can
exponentially blow up, as indicated by numerical simulations.  Since
the chaotic orbits are more sensitive to perturbations than regular
(periodic) ones, external/internal forcing are more effective on this
chaotic component. In this paper, we investigate chaotic and complex
phenomena related to the formation of a central bar which give rise to
the diffusion of the stellar component in the disk.

The bar has a time-dependent activity, with a pattern speed which
typically decreases in isolated galaxies (Sellwood~\cite{sellwood}). However,
the system can be cooled or heated by energy dissipation or infall of
gas, or by forming stars on low-velocity dispersion orbits, with the
net effect of impacting the amplitude of spiral waves and the strength
of the bar, or even destroying it. In this way bars (and spiral waves)
can be seen as recurrent patterns which can be rebuilt during their
long history until the present configuration at redshift $z=0$
(Bournaud \& Combes~\cite{bournaud}). Under the action of these non-axisymmetric
patterns, stars move in the disk and gradually increase their velocity
dispersion, as suggested by observations in the Solar neighborhood
(see Holmberg et al.~\cite{holmberg09} and references therein) and in external
galaxies (Gerssen et al.~\cite{gerssen}, Shapiro et al.~\cite{shapiro}). 
The origin and the amount of disk heating are still open to debate.

First attempts to explain such a heating process in disk galaxies were
made by empirically modeling the observed increase of the stellar
velocity dispersion with age in the solar neighborhood.  
Wielen~(\cite{wielen})
suggested a diffusion mechanism in velocity space, which gives rise to
typical relaxation times for young disk stars of the order of the
period of revolution and to a deviation of stellar positions of
1.5\,kpc in $200$\,Myr. The result was obtained without making
detailed assumptions on the underlying local acceleration process
responsible for the diffusion of stellar orbits. Global acceleration
processes, such as the gravitational field of stationary density waves
or of central bars with constant pattern speed, were ruled out since
their contribution to the velocity dispersion of old stars was 
found to be negligible and concentrated in particular resonance
regions (Wielen~\cite{wielen}; 
Binney \& Tremaine~\cite{binneytremaine}, p.~693). 
In isolated galaxies, different local accelerating mechanisms have been
investigated, such as the gravitational encounters between stars and
giant molecular clouds (Spitzer \& Schwarschild~\cite{spitzer51,spitzer53}; 
Lacey~\cite{lacey}), secular heating produced by transient spiral arms 
(Barbanis \& Woltjer~\cite{barbanis}; Carlberg \& Sellwood~\cite{carlberg}; 
Fuchs~\cite{fuchs}) or the
combination of the two processes (Binney \& Lacey~\cite{binney88}; Jenkins \&
Binney~\cite{jenkins}). Another heating mechanism was suggested by Minchev and
Quillen~(\cite{minchev06}), who showed that the stellar velocity dispersion can
increase with time due to the non-linear coupling between two spiral
density waves.

Such local acceleration mechanisms suggest the existence of a
significant component of the galactic gravitational field with a
rather chaotic behavior. Pfenniger~(\cite{pfenniger86}) investigated the relation
between diffusion and chaotic orbits. The latter typically react
promptly to small perturbations. He pointed out that the effect of
perturbations on regular orbits, such as epicyclic orbits,
underestimates strongly the stellar diffusion rate when a stellar
system becomes non-integrable, as for example in the presence of a
central bar. As the central bar develops, reaches its maximal
amplitude and then settles down to an almost steady state, its
gravitational potential changes in time. In time-dependent potentials,
the number of chaotic orbits typically decreases 
while the system secularly evolves toward a quasi-steady state
through collective effects. The system is then ready again to respond
(mainly through the remaining irregular orbits) to external
perturbations, such as new infall of gas, and to recurrently restore a
strong bar. The bar is thus able to perturb orbits of stars born or
passing through its region, which can visit at later times the Solar
neighborhood (Raboud et al.\ \cite{raboud}). Indeed, most of the observational
signatures of radial mixing reported in the literature 
(Grenon~\cite{grenon72,grenon99}; Castro et al.\ \cite{castro}) 
point to stars coming from a region next to
the bulge/bar intersection, suggesting the bar to be a key player in
the radial migration process.

The subject of radial migration of stars was revived when a large
scatter in the observed age-metallicity relation (AMR) was reported by
Edvardsson et al.\ (\cite{edvardsson}), later confirmed by the larger
Geneva-Copenhagen Survey sample (Nordstr\"om et al.\ \cite{nordstrom}, 
Holmberg et al.\ \cite{holmberg07,holmberg09}; see also Casagrande et 
al.\ \cite{casagrande}). It must be said
that even though the AMR has been extensively studied in the solar
vicinity, the results are still controversial due to the large
uncertainties in stellar ages (see Pont \& Eyer~\cite{pont}). For instance,
using a sample of stars for which it was possible to obtain
chromospheric ages, Rocha-Pinto et al.\ (\cite{rocha}) have reported a much 
tighter AMR.

The mechanisms driving radial diffusion and heating are still hotly
debated, and in many cases the role of the bar is not taken into
account. Assuming the whole scatter seen in Edvardsson et al. 
(\cite{edvardsson}) data was real, 
Sellwood \& Binney (\cite{sellwoodbinney}) pointed out that the radial
excursion predicted by Wielen (\cite{wielen}) was not sufficient to explain the
weakness of the AMR in the solar neighborhood.  In order to explain
both the large scatter in the AMR and the evidence that even old disk
stars today have nearly circular orbits, Sellwood \& Binney 
(\cite{sellwoodbinney})
suggested a new mechanism based on the resonant scattering of stars
under the effect of transient spiral waves.  In this process, a star
initially on a nearly circular orbit resonates with a rotating wave
and changes its angular momentum. If the duration of the peak
amplitude of the perturbing potential is less than the period of the
`horseshoe' orbits, i.e.~orbits of particles trapped at the corotation
radius of the spiral wave, the star can escape from the potential well
without changing its eccentricity.  The net effect of this scattering
mechanism is that stars migrate radially without heating the disk. In
other words, the overall distribution of angular momentum is
preserved, except near the corotation region of the transient spiral
wave, where stars can have large changes of their angular
momenta. Haywood (\cite{haywood}) estimated upper values for the migration rate
from 1.5 to 3.7~kpc/Gyr, which agree with the values in L\`epine et
al.\ (\cite{lepine}) for the radial wandering due to the scattering mechanism
assumed by Sellwood \& Binney (\cite{sellwoodbinney}).

Radial diffusion of stars (and gas) could have important implications
for the interpretation of key observational constraints for the 
formation of the Galaxy, such as the
AMR, metallicity distributions, or the metallicity gradients, since
old, probably more metal rich stars that formed at small
galactocentric radii, as well as young metal-poor stars formed at
large radii are enabled to appear in Solar-neighborhood samples 
(e.g. Haywood~\cite{haywood}). Due
to the lack of detailed information on the processes driving stellar
radial migration, models of the Galactic chemical evolution have
evaluated past history of the solar neighborhood and the formation and
evolution of the abundance gradients assuming that the Galaxy can be
divided into concentric wide ($\sim$1-2~kpc) cylindrical annuli, which
evolve independently (van den Bergh~\cite{vandenbergh}; Schmidt~\cite{schmidt}; 
Pagel~\cite{pagel}; Chiappini et al.\ \cite{chiappini97}, 
Chiappini et al.\ \cite{chiappini01}).  Sch\"onrich \&
Binney (\cite{schonrich09a}) explored the consequences of mass exchanges between
annuli by taking into account the effect of the resonant scattering of
stars described before.  This approach appears to be successful to
replicate many properties of the thick disk in the Solar neighborhood
without requiring any merger or tidal event (Sch\"onrich \& Binney
\cite{schonrich09b}). 
High resolution cosmological simulations (Ro\v{s}kar et al.\ 
\cite{roskar}, Loebman et al.\ \cite{loebman}) 
give support to the view that such
scattering mechanism determines a significant migration in the stellar
disk. However, the strong mixing driven by bar resonances was not
taken into account (see below), casting thus doubts on some of the
conclusions in the papers quoted above.

The Milky Way (MW) is a barred galaxy and it is clear that the process
above, not accounting for the existence of the bar, is probably just
one of the processes at play among others. Indeed, the role of
resonant couplings between bars and spirals (Tagger et al.\ \cite{tagger}) in
the distribution of energy and angular momentum in disk galaxies could
also play a major role.  Recently, Minchev and collaborators (Minchev
\& Famaey~\cite{minchev10}, Minchev et al.\ \cite{minchev11}) 
have further analyzed this
mixing mechanism finding that resonances between the bar and the
spiral arms can act much more efficiently than transient spiral
structures, dramatically reducing the predicted mixing time-scales. 
Moreover, while for the Sellwood \& Binney (\cite{sellwoodbinney}) mechanism to
work short-lived transient spirals are required, in barred galaxies,
such as the MW, spirals are most likely coupled
with the bar as shown by Sparke \& Sellwood~\cite{sparke},
and thus longer lived (Binney \& Tremaine~\cite{binneytremaine}, Quillen
et al.\ \cite{quillen}). 
As a consequence the radial migration process in the MW could
have been different than currently predicted.

In order to include the effect of radial migration in chemical
evolution models and to gain a global (chemical and kinematic)
understanding of the processes at play in the galactic disks, many
dynamical aspects need to be further investigated and in particular
the role of the bar, that is the strongest non-axisymmetric component
in disk galaxies. In this paper, we present $N$-body simulations of
barred spiral galaxies, and study how disks with different degrees of
stability, ranging from marginally stable disks with Safronov-Toomre
parameter $Q_T \sim 1$ to hot disks with $Q_T > 1$, respond to the
presence of bar patterns. Our aim is to estimate the time and length
scales of stellar diffusion in the radial direction and to relate
these quantities to the strength of the bar and to the number of hot
particles in the disk, i.e. generally chaotic particles which are
susceptible to cross the corotation barrier and to explore all space,
being characterized by values of the Jacobi integral $H$ larger than
the value at the Lagrangian points, $H > H(L_{1,2})$ (Sparke \&
Sellwood~\cite{sparke}; Pfenniger \& Friedli~\cite{pfenniger91}).  
We investigate how these
characteristic scales evolve in time, and depend on the activity of
the bar. We consider different scenarios of diffusion, and discuss
their implications for chemical evolution constraints in our Galaxy.

The paper is organized as follows. In Sect.~2 we describe the
simulations and the relevant parameters. In Sect.~3 we solve the
diffusion equation in axisymmetric systems, we define the diffusion
coefficient, the diffusion time-scale and the diffusion length-scale,
and the methods used to estimate these quantities from the simulation
results. In Sect.~4 we present our results. In Sect.~5 we discuss the
implications for chemical evolution models of the MW and we summarise
our findings.

\section{$N$-body simulations}

We have run self-consistent $N$-body simulations starting from a
bar-unstable axisymmetric model. We have analyzed initial
configurations with disk, bulge and dark halo components which differ
on the initial value of the Safronov-Toomre parameter $Q_T =
\sigma_r\, \kappa/ (3.36\, G\, \Sigma)$ (Safronov~\cite{safronov}; 
Toomre~\cite{toomre}),
where $\sigma_r$ is the radial velocity dispersion of the disk
component, $G$ is the gravitational constant, $\Sigma$ is the disk
surface density, $\kappa$ is the epicycle frequency defined by
$\kappa^2 = R d\Omega^2/dR + 4\Omega^2$, where $\Omega$ is the
circular frequency related to the global gravitational potential
$\Phi(R,z,t)$ in the disk plane $z=0$ by $\Omega^2 = (1/R)\, \partial
\Phi/\partial R$.

The initial mass distribution in our simulations corresponds to a
superposition of a pair of axisymmetric Miyamoto-Nagai disks of mass
$M_B$, $M_D$, horizontal scales $A_B+B$, $A_D+B$, and identical
scale-height $B$,
\begin{equation}\label{MN}
  \Phi_{MN}(R,z)=\sum_{i=B,D} \frac{-GM_i}{\sqrt{R^2+(A_i+\sqrt{B^2+z^2})^2}}
\end{equation}
The first component represents the bulge ($B$), while the second the
disk ($D$) (Pfenniger \& Friedli~\cite{pfenniger91}). 
The parameters have been set
to $A_B = 0.07$~kpc, $A_D = 1.5$~kpc, $B=0.5$~kpc, $M_B/M_D =
3/17$. The initial particle positions and velocities are found by a
Monte-Carlo draw following the density law corresponding to
Eq.~(\ref{MN}), truncated to a spheroid of semi-axes $R=30$~kpc,
$z=10$~kpc.  The number of particles in the disk-bulge component is $N
= 4\cdot 10^6$ and the total mass is $M_{BD} = 4.2\cdot
10^{10}~M_\odot$.

In order to progressively heat the disk, we have added to this
bulge-disk component an oblate pseudo-isothermal halo with the
following density distribution (except in models {\tt m1} and {\tt
  m2}):
\begin{equation}\label{halo}
\rho_H(R,z)= \frac{\rho_h}{1+R^2/R_h^2+z^2/z_h^2}
\end{equation}
The number of particles in this halo component is $N_{H}=2\cdot 10^6$,
which is a value in the range suggested in Dubinski et al.\ \cite{dubinski}) in
order to obtain convergent behavior in studies of bar formation and
evolution.  The length-scales are $R_h = 7.5$~kpc and $z_h =
3.5$~kpc. The density distribution has been truncated to $R=30$~kpc,
$z=15$~kpc. We set the total mass in the dark halo $M_H$ to be four
times the total mass in the bulge-disk component $M_{BD}$, except in
the model {\tt m3}, where $M_H = 2 M_{BD}$ (see Table~\ref{table:1}).
The effect of adding the halo component is that the
bar becomes progressively smaller and with higher pattern speed, the
disk is hotter and less sensitive to the bar perturbations.

We then impose the equilibrium of the first and second moments of
velocities by solving Jeans' equations (see e.g., Binney \& 
Tremaine~\cite{binneytremaine}) 
with a constant $Q_T$.  The resulting distribution is relaxed
for a couple of rotations until ripples spreading through the disk
from the center disappear.  We use this as the initial condition for
the $N$-body simulations performed by using the {\tt Gadget-2}
free source code (Springel et al.\ \cite{springel01}, 
Springel~\cite{springel05}).

\begin{figure*}
   \centering
   \includegraphics[width=6cm]{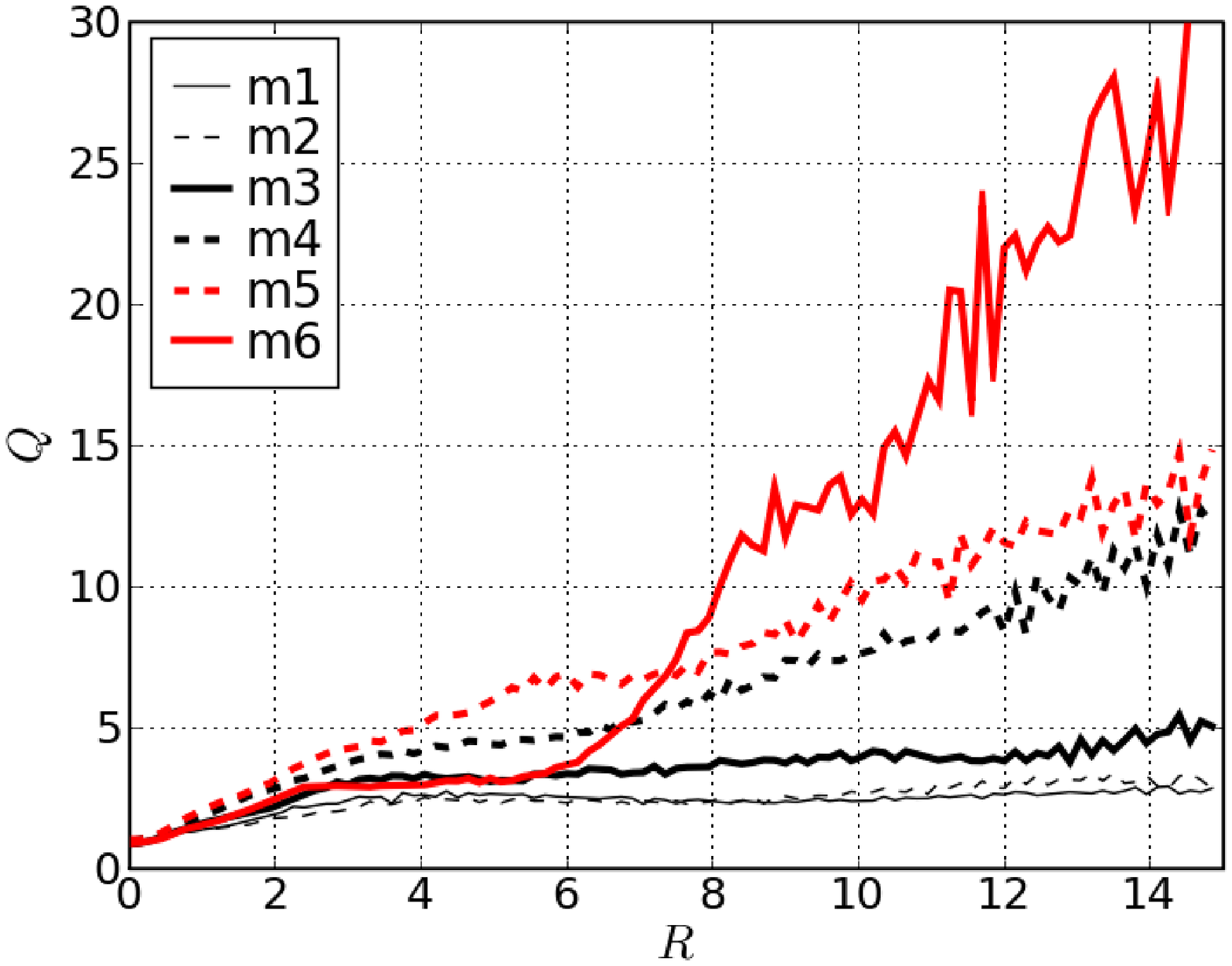}
   \includegraphics[width=6cm]{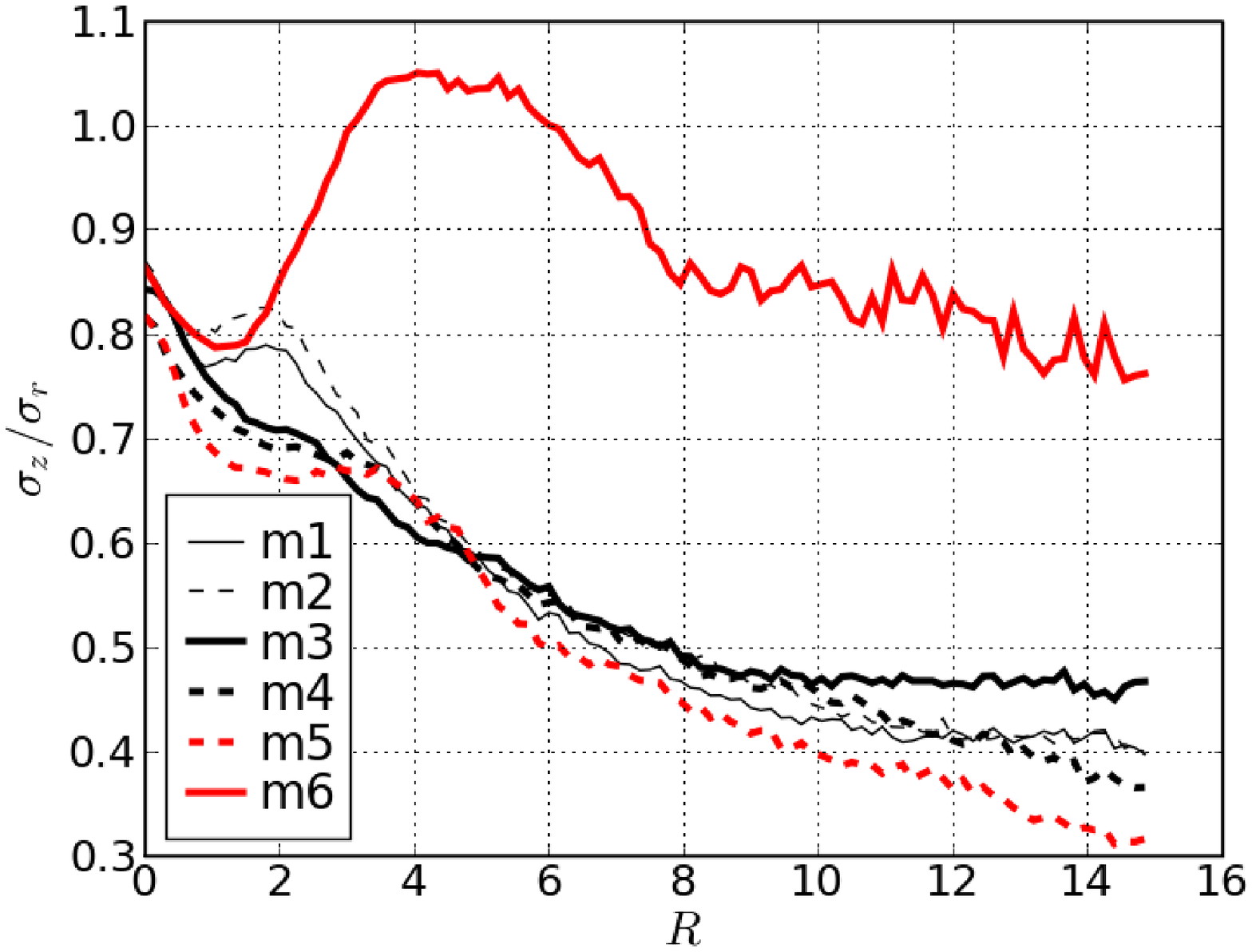}
   \includegraphics[width=6cm]{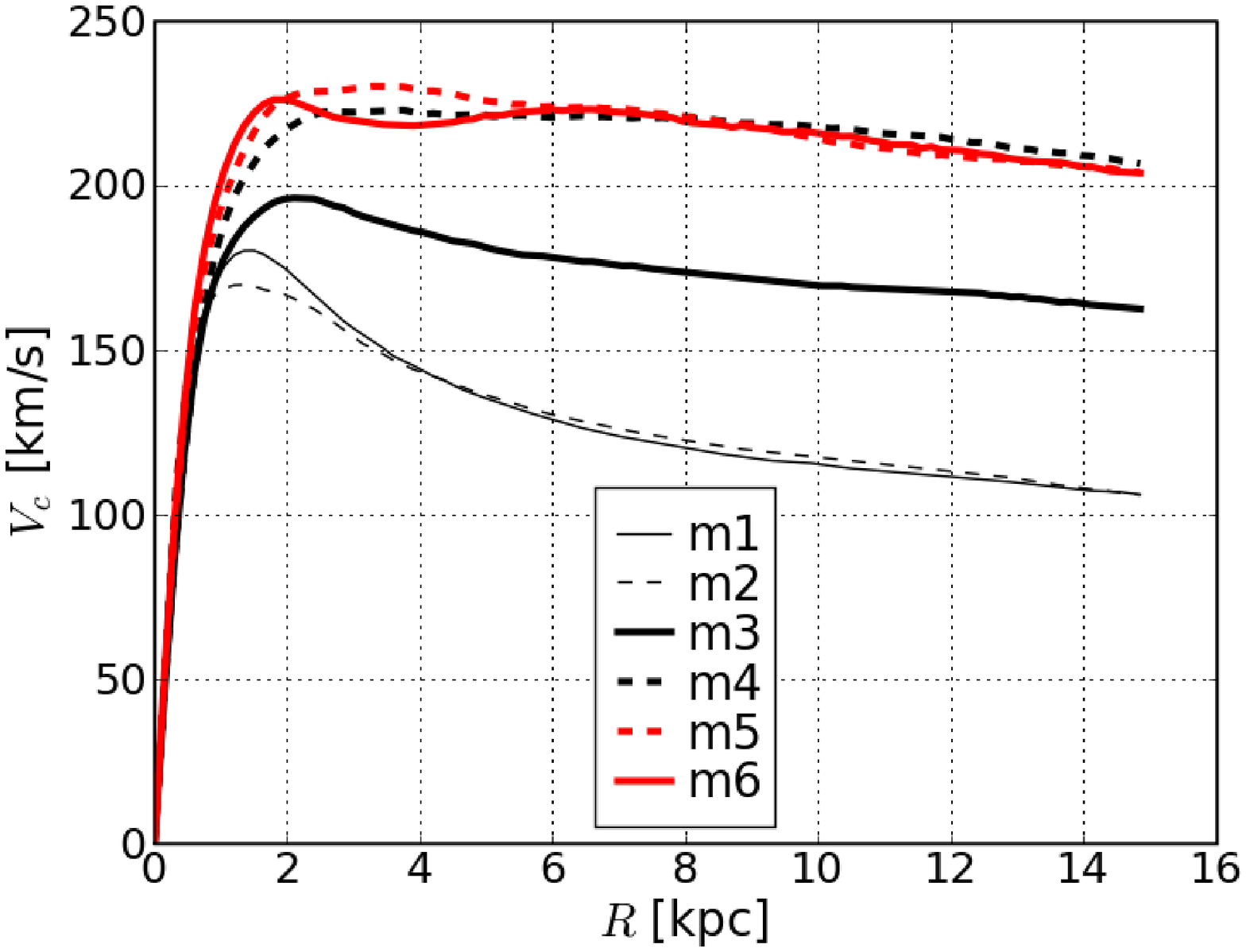}
   \caption{\small {\it Left}:
   Safronov-Toomre-parameter at the final time. {\it Middle}: Araki parameter at
   the final time. {\it Right}: rotation curves.}
   \label{RotCurve}%
\end{figure*}

The initial {\tt Gadget-2} configurations considered in this work
differ on the values of the Safronov-Toomre parameter, which ranges
from $Q_T \sim 5$ at two scale lengths from the center for hot disks
to $Q_T \sim 1$ for marginally stable disks. These $Q_T$ values are
listed in Table~\ref{table:1}, along with the initial values of the
radial and vertical velocity dispersions at two scale lengths from the
center. We have considered these initial values in order to
investigate how the radial diffusion depends on the disk sensitivity
to perturbations. Thus, we have considered two extreme cases: in one
case the disk is marginally stable and spiral waves develop (models
{\tt m1} and {\tt m2}), with the main global effect of heating in the
radial direction (the Araki parameter $\sigma_z/\sigma_r \sim 0.5$ at
two scale lengths from the center at the end of the simulations and it
decreases at larger radii, see the middle panel of
Fig.~\ref{RotCurve}). In the other case (model {\tt m6}), the disk is
heated and the velocity dispersions increase in both the radial and
the vertical direction (the Araki parameter $\sigma_z/\sigma_r \sim
0.8$ at large radii at the end of the simulation, see the
corresponding curve in the middle panel of Fig.~\ref{RotCurve}). The
other models considered have intermediate values of $Q_T$ and
$\sigma_z/\sigma_r$ between these two extreme cases. The
Safronov-Toomre and Araki parameters at the final times are shown in
the left and middle panels, respectively, of Fig.~\ref{RotCurve}.

\begin{table}
\caption{Initial configuration of the $N$-body simulations: name of
  the model, initial Safronov-Toomre parameter at two scale lengths from
the center, ratio between halo and disk-bulge masses, initial radial and
  vertical velocity dispersions at two scale lengths from
the center, ratio between hot particles and total visible particles.}
\label{table:1}
\centering
\begin{tabular}{clcccr}
\hline\hline
Model & $Q_T$ & $M_H/M_{BD}$ & $\sigma_r$ [km/s] & $\sigma_z$ [km/s] &
$N_{hot}/N$ \\
\hline
{\tt m1}    &  1      &  {\bf --}  & 20 & 18 & 12\%  \\
{\tt m2}    &  1.5      &  {\bf --}  & 30 & 20 & 15\%  \\
{\tt m3}    &  2      &  2  & 30 & 20 & 4\%   \\
{\tt m4}    &  4      &  4  & 30 & 22 & 2.5\% \\
{\tt m5}    &  5      &  4  & 40 & 22 & 2.5\%     \\
{\tt m6}    &  5      &  4  & 40 & 40 & 5\%   \\ 
\hline
\end{tabular}
\end{table}
 
In the right panel of Fig.~\ref{RotCurve} the rotation curves for all
the $N$-body models listed in Table~1 are shown at the final time.
When the halo component is included (in models {\tt m3, m4, m5, m6}),
the rotation curve is flatter and $V_{\rm c} \sim
220~\textrm{km}~\textrm{s}^{-1}$ at two scale lengths from the center,
that is in the region between 6 and 10~kpc, depending on the
model. The rotation curve of model {\tt m3}, where the ratio between
halo and disk-bulge mass is $M_H/M_{BD} = 2$, has intermediate values
between that of models without the halo ({\tt m1, m2}) and the others.

We classify stellar orbits into three dynamical categories (Sparke \&
Sellwood~\cite{sparke}; Pfenniger \& Friedli~\cite{pfenniger91}). 
The first two dynamical
categories are the bar and the disk orbits with the Jacobi integral
$H=E-\Omega_p L_z$ smaller than the value at the Lagrangian points
$L_{1,2}$, $H< H(L_{1,2})$, where $E$ is the total energy and $L_z$ is
the $z$-component of the angular momentum. The separation of particles
in the bar or disk component can be easily done since bar orbits
typically have smaller values of $L_z$ and $E$ than disk orbits. The
third category includes hot orbits for which $H\ge H(L_{1,2})$.  The
models considered here differ in the number of hot particles after the
formation of the bar (the ratio between the number of stars in the hot
component and the total number of the visible stars is listed in the
last column of Table~\ref{table:1}).

The bar pattern speed $\Omega_P \equiv \frac{d\theta}{dt}(t)$, where
$\theta$ is the azimuthal angle of the bar major axis (in the inertial
frame) calculated by diagonalising the moment of inertia tensor of the
bar particles, ranges from
$35~\textrm{km}~\textrm{s}^{-1}~\textrm{kpc}^{-1}$ for model {\tt m1}
to $40~\textrm{km}~\textrm{s}^{-1}~\textrm{kpc}^{-1}$ for {\tt m6}, at
final times.  These values are comparable to those found by Fux (\cite{fux}
- see his Fig.~5), while recent estimates of the MW bar pattern speed are 
of the order of $50-60~\textrm{km}~\textrm{s}^{-1}~\textrm{kpc}^{-1}$ (Dehnen~\cite{dehnen2000}, Minchev et al.~\cite{minchev2007,minchev2010}).  
The pattern speed is typically slowly decreasing in time with a rate of a few km/s/kpc/Gyr
(Fux~\cite{fux}; Bournaud \& Combes~\cite{bournaud}), so that the values at the
beginning of the simulations are
$60~\textrm{km}~\textrm{s}^{-1}~\textrm{kpc}^{-1}$ and
$80~\textrm{km}~\textrm{s}^{-1}~\textrm{kpc}^{-1}$ for models {\tt m1}
and {\tt m6}, respectively.  The corresponding corotation radius
$R_c$, obtained by the intersection of $\Omega_P$ with the circular
frequency, $\Omega_P(t) = \Omega(R_c,t)$, increases in time (it
typically ranges from $R_c = 2$ to 5~kpc at final times in our
models).

\begin{figure}
   \centering
   \includegraphics[width=7cm]{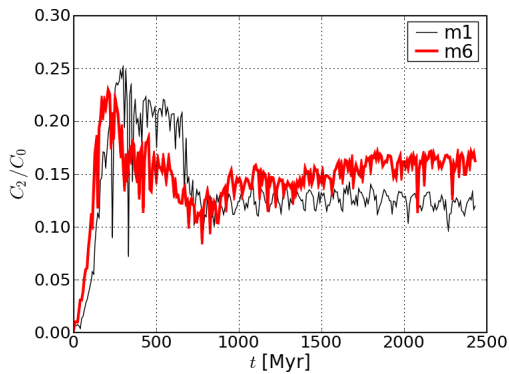}
   \caption{\small Evolution in time of the bar's strength for models
   {\tt m1} and {\tt m6}.}
   \label{figC2Bar}%
\end{figure}

In order to understand the role played by the central bar on the
distribution of stars in the disk, we follow the evolution of the
strength of the bar in time for each model.  If $C_m$ is the amplitude
of the mode $m$ in the density distribution,
\begin{equation}
C_{m} = \left| \sum_{j} \exp(i\, m\theta_j)\right|,\quad
\end{equation}
the bar's strength is defined as the mode $C_{2}$ when the stars $j$
are restricted to the bar component.  We normalise $C_2$ with respect
to the number of stars in the bar component, $C_0$. This quantity is
shown in Fig.~\ref{figC2Bar} for models {\tt m1} and {\tt m6}.  Since
we do not include the gas component in our models, we are not able to
follow their evolution for times longer than a few
Gyrs. After this typical time-scale, the bar
amplitude saturates and the systems reach a quasi-steady state. Since
the inclusion of the gas component necessarily requires the
introduction of other less controlled parameters, such as the cooling
rate or star formation, we prefer in this first study to limit the
integration time over a couple of Gyrs, which is
already enough to study the role of the bar in the radial migration
process.

\begin{figure*}
  \centering
   \includegraphics[width=7cm]{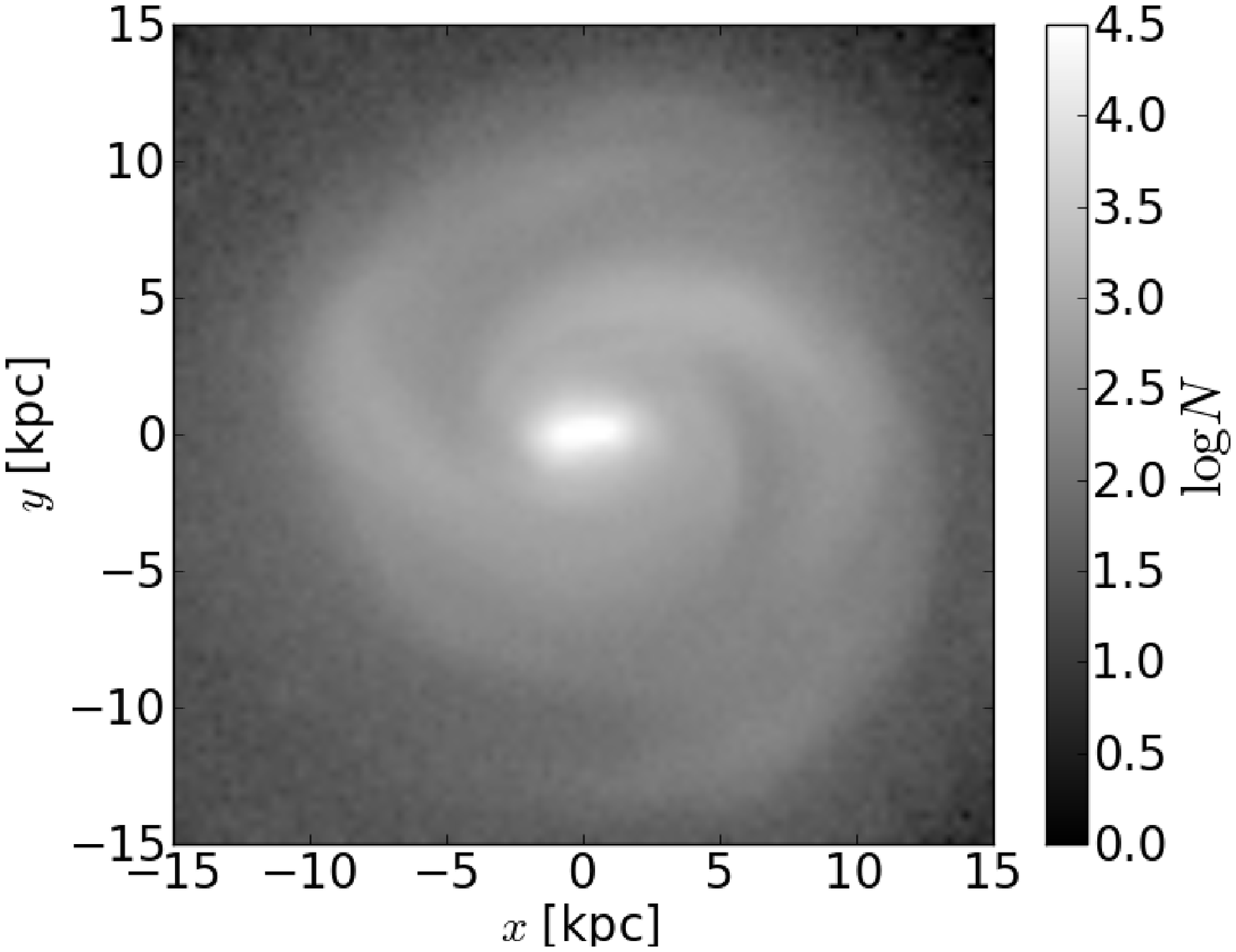}
   \includegraphics[width=7cm]{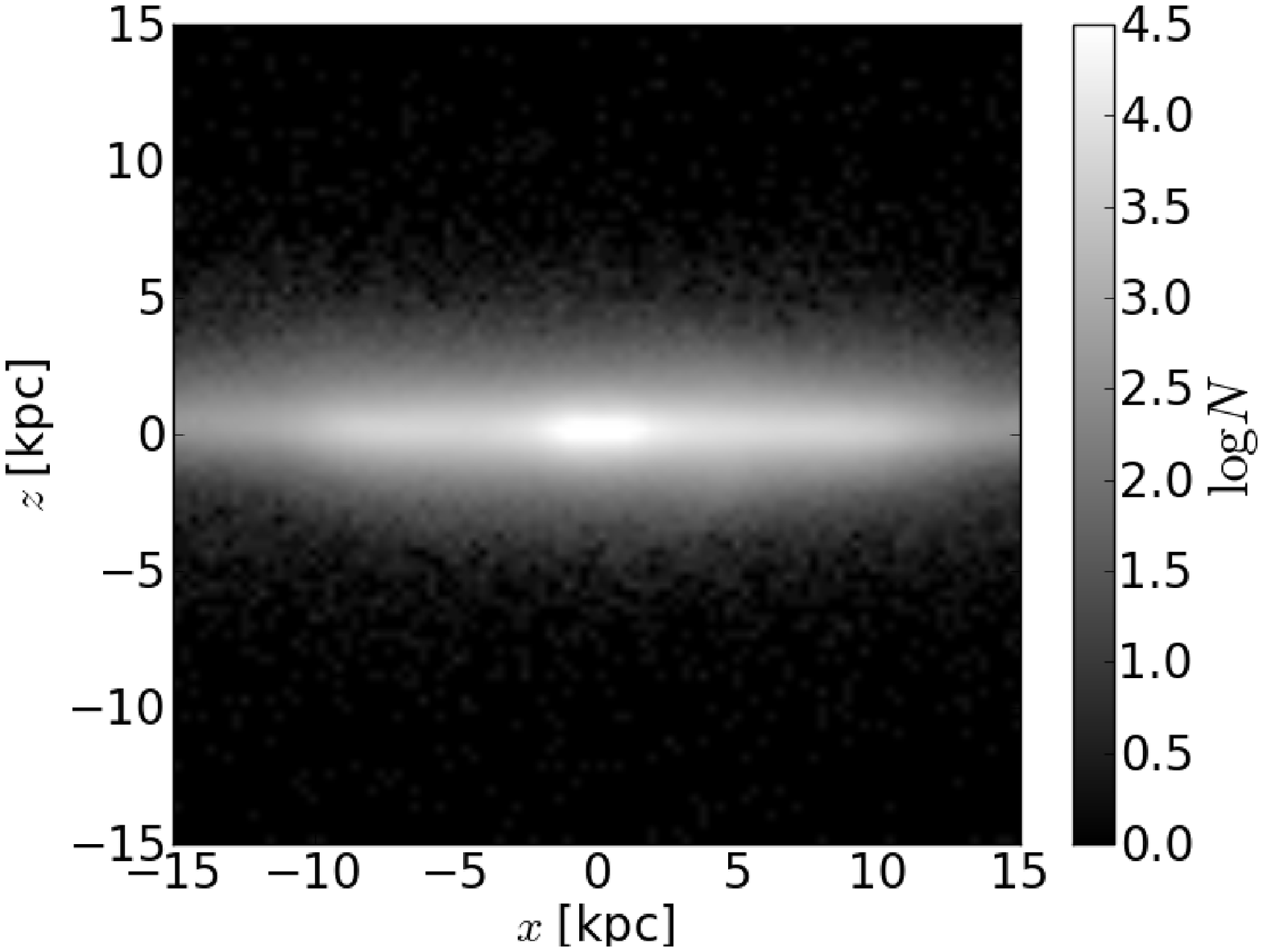} 
   \includegraphics[width=7cm]{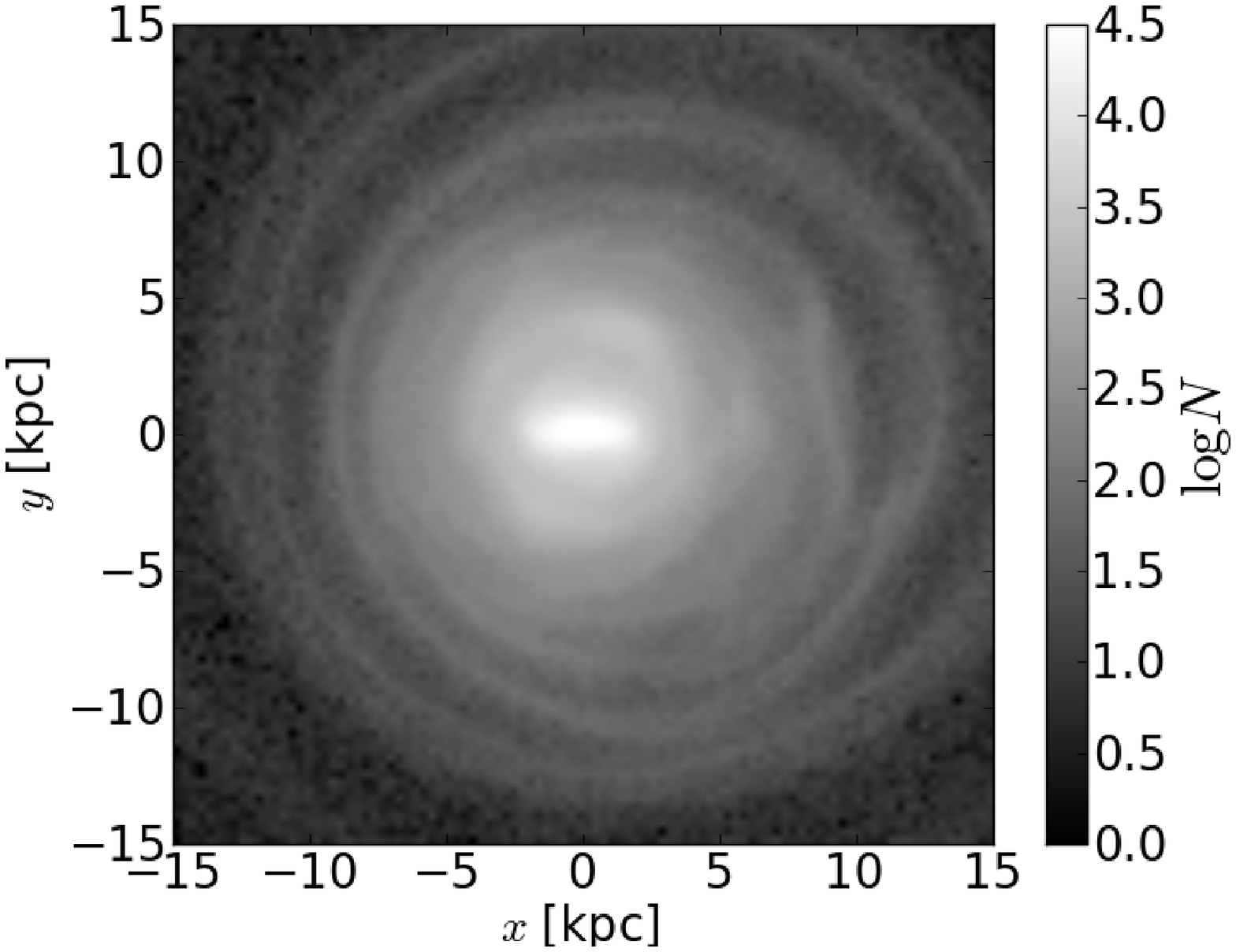}
   \includegraphics[width=7cm]{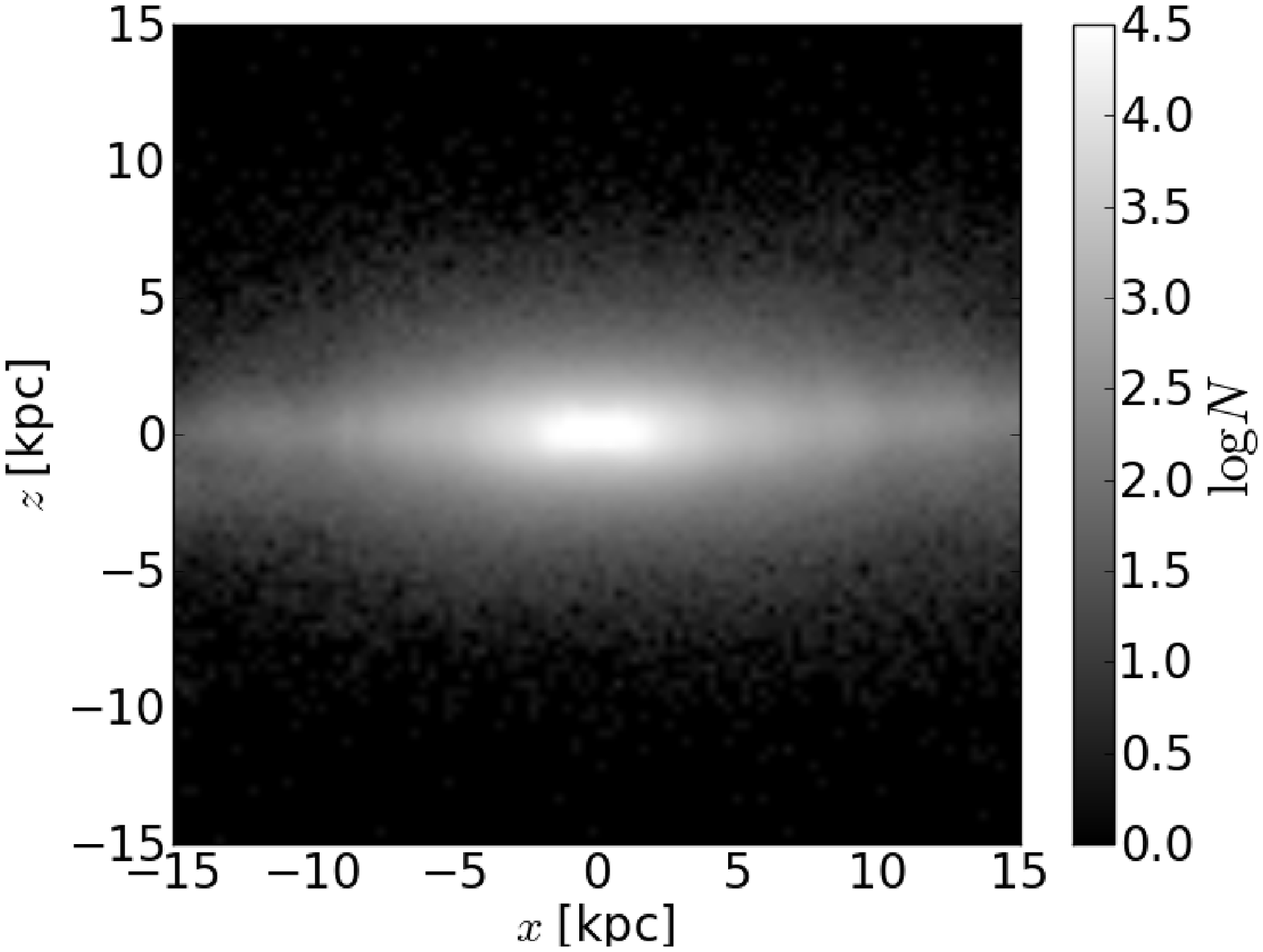} 
   \caption{Density maps at $t\sim 550$~Myr. {\it Right}: face-on
   views, {\it left}: edge-on views. {\it Top panels}: model {\tt m1},
   {\it bottom panels}: model {\tt m6}.}
   \label{figDensity}
\end{figure*}

The face-on and edge-on views of the density distribution of models
{\tt m1} (upper panels) and {\tt m6} (bottom pannels) are shown in
Fig.~\ref{figDensity} at time $t \sim 550$~Myr, that is just after the
maximum strength of the bar
(cf. Fig.~\ref{figC2Bar}). In the first case, both
bar and spiral arms develop since the disk is sufficiently cold, while
in the second case the disk is hot and only a bar (with a smaller
corotation radius than in the previous case) develops in the central
region.  
In the external regions of models {\tt m1} and {\tt m6}, 
other patterns can be observed, the dominant being the
pattern with $m =1$. This mode is related to asymmetric 
distributions of mass pushed by the system rotation and 
it can give rise to filaments of stars or ring-like structures on 
long time-scales.

\section{Diffusion equation in axisymmetric systems}
 
In this paper we model the diffusion along the
radial direction in the galactic plane by introducing a
distribution function $F(R,t)$ which satisfies the phenomenological
{\it spatial} diffusion equation (from Fourier's law) in cylindrical 
coordinates:
\begin{equation}
  \partial_t F = \frac{1}{R}\, \partial_R (R D \partial_R F)\, ,
\label{DiffEq}
\end{equation}
where $D$ is the diffusion coefficient (in unit of area per time).
This diffusion model has already been more succinctly presented in
Brunetti et al.\ (\cite{brunetti}). We suppose that for a limited time all the
stochastic processes ongoing in a spiral galaxy can be described by
such an equation characterized locally by a single positive diffusion
coefficient $D$, to be empirically measured in the full dynamical
simulations for a range of $R$ and $t$.  The parameter $D$ conveys the
notion that the diffusion process is a surface increase per unit time.
For the sake of simplicity we choose $D$ to be constant and
independent of $F$, which makes Eq.~(\ref{DiffEq}) a linear partial
differential equation for $F$.  Another choice could have been to take
$C \equiv D \rho$ as a local parameter, where $\rho$ is the mass density.
The unit of $C$ is mass per length per time which conveys then the
notion of a mass flux gradient. However this choice would have led to
a more difficult empirical determination of $C$ since the equation
involves then $\rho$ and its gradient. Thus, the model considered here gives a lower limit on the diffusion happening in a barred galaxy, since other more complex diffusion processes are neglected in the present analysis.

\begin{figure}
   \centering
   \includegraphics[width=7cm]{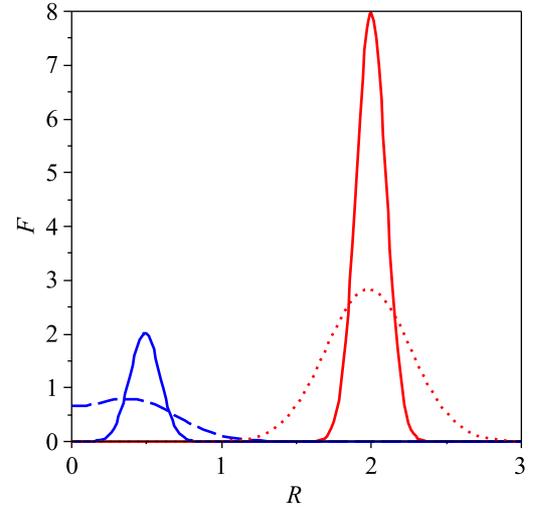}
     \caption{\small Two distributions with $D = 1$ and centered in
   $R_0 = 0.5$ and 2 (solid lines) evolve in time (dashed and dotted
   lines, respectively), as described by
   Eq.~(\ref{eqDiff}). }
    \label{fig:solution}
\end{figure}

The general solution of Eq.~(\ref{DiffEq}) with constant $D$, which is
non-singular at $R=0$ is given by:
\begin{equation}
  F(R,t) = \int_0^\infty A(s)\, e^{-D t s^2} J_0(sR)\,s\, ds\, ,
\label{sol1}
\end{equation}
where $J_0(x) = J_0(-x)$ is the Bessel function of the first kind.
The function $A$ can be determined by taking the Hankel transform of
$F(R,0)$:
\begin{equation}\label{Hankel}
  A(s) = \int_0^\infty F(R,0)\, J_0(sR)\, R\, dR\, 
\end{equation}
By inserting Eq.~(\ref{Hankel}) into Eq.~(\ref{sol1}) and assuming
that the particles are initially localized at a certain radius $R_0$
at time $t_0 =0$, $F(R,0) = F_0R_0\delta(R-R_0)$, we obtain:
\begin{equation}
  F(R,t) = R_0^2 F_0 \int_0^\infty s\, e^{-D t s^2} J_0(sR)\, J_0(sR_0)\,
  ds\, 
\end{equation}
Thus, the diffusion of this distribution can be expressed in terms of
Bessel and elementary functions (Gradsteyn \& Ryzhik~\cite{gradsteyn}, formula
6.633.2).  The time-evolution of an initial set of localized particles
reads:
\begin{equation}\label{eqDiff}
  F(R,t) = \frac{R_0^2 F_0}{2D t} 
  \exp\left(-\frac{R_0^2+R^2}{4D t}\right)\,
  I_0\left(\frac{RR_0}{2D t}\right)\, , 
\end{equation}
where $I_0(x)$ is the modified Bessel function of the first kind,
which is finite at the origin, $I_0(0)=1$.  In Fig.~\ref{fig:solution}
two initial distributions with $D = 1$ and centered in $R_0 = 0.5$ and
2 (solid lines) evolve in time, as described by Eq.~(\ref{eqDiff})
(dashed and dotted lines, respectively). At large radii, the
distributions are essentially Gaussian, while at small radii they are
strongly modified from the contributions of particles at the center of
the cylinder.  Eq.~(\ref{eqDiff}) describes the distribution of the
radial positions of stars in the disk at the initial time $t_i$ which
diffuse toward position $R_0$ at time $t_0$, such that $t_0 - t_i =
\Delta t \le T_D$, where $T_D$ is the diffusion time-scale or,
equivalently, the distribution of stars which initially are in $R_0$
at $t_0$ and then diffuse toward $R$ with $\Delta t \le T_D$.

When $R$ goes to zero, Eq.~(\ref{eqDiff}) reduces to:
\begin{equation}
  F(0,t) = \frac{R_0^2 F_0}{2D t} \, \exp\left[-R_0^2/(4D
    t)\right]\, 
\end{equation} 
For large values of the argument the modified Bessel function $I_0(x)
\to (2\pi x)^{-1/2}\, \exp(x)$ and thus $F(R,t)$ reduces to:
\begin{eqnarray}
  \lim_{R\to \infty} F(R,t) &=& \frac{R_0^{3/2} F_0}{\sqrt{4\pi D t R}} 
  \exp\left(-\frac{(R-R_0)^2}{4D t}\right) \nonumber \\
  &=&  \frac{R_0^{3/2}
    F_0}{\sqrt{R}}\, {\cal{N}}(\mu,\sigma) \label{limEqDiff}
\end{eqnarray}
where ${\cal N}(\mu, \sigma)$ is the Gaussian distribution with mean
value $\mu = \langle R\rangle = R_0$ and standard deviation $\sigma =
\sqrt{2D t}$.

In order to obtain a simple model of the distribution of the stars in
a galactic disk, one can consider to envelop Eq.~(\ref{limEqDiff}) by
an exponential surface density $\Sigma(R) \propto \exp(-R/R_d)$ (see,
for example, Sellwood \& Binney~\cite{sellwoodbinney}), 
where $R_d$ is the disk scale length, thus obtaining:
\begin{eqnarray}
  p_d(R,t) &=& \frac{R_0^{3/2} F_0}{\sqrt{4\pi D t R}}\, C\,  
  \exp\left(-\frac{(R-R_0)^2}{4D t}\right)\, \exp(-R/R_d) \nonumber \\
  &=& \frac{C' \sqrt{R_0}}{\sqrt{4\pi D t R}}
  \exp\left(-\frac{(R-R_0+\sigma^2/R_d)^2}{4D t}\right)\,  \nonumber \\
  &=& C' \sqrt{\frac{R_0}{R}}\,  {\cal N}(\mu', \sigma) \label{DiskProb}
\end{eqnarray}
where $C$ and $C'$ are normalization constants and ${\cal N}(\mu',
\sigma)$ is the Gaussian distribution with mean value $\mu' = \langle
R\rangle = R_0 - \sigma^2/R_d$ and standard deviation $\sigma =
\sqrt{2D t}$. It is important to remember that the diffusion model
used for obtaining Eq.~(\ref{limEqDiff}) is only valid for times
smaller than the diffusion time-scale.
 
We set $R_0 = R_\odot = 8$~kpc and $R_d \sim 3$~kpc. We consider the
distribution of stars in $R \sim R_\odot$ at the initial time.  The
previous expression, Eq.~(\ref{DiskProb}), can be used to estimate the
relative fraction of stars which remain in a diffusion time-scale
within the local volume $|R-R_\odot|\le d$. This is given by:
\begin{equation}
  p(|R-R_\odot| \le d) = 
  \int_{R_\odot-d}^{R_\odot+d} {\cal N}(\mu', \sigma)\, dR 
\end{equation} 
The integral can be written as:  
\begin{equation}\label{ErrF}
  p(|R-R_\odot| \le d) = 
  \frac{1}{\sqrt{\pi}} \int_{x_-}^{x_+} e^{-x^2}\, dx = 
  \frac{1}{2} [\textrm{erf}(x_+)- \textrm{erf}(x_-)]
\label{LocalVolume}
\end{equation} 
where $x_\pm = (\sigma/R_d \pm d/\sigma)/\sqrt{2}$. If $d\ll \sigma$,
we get $x_+ \sim x_-$ and thus the probability of staying in the local
volume is nearly zero.  If $d \sim \sigma \ll R_d$, we have
$p(|R-R_\odot| \le d) \sim \frac{1}{2} [\textrm{erf}(1/\sqrt{2})-
\textrm{erf}(-1/\sqrt{2})] = \textrm{erf}(1/\sqrt{2}) = 0.68$.  If $d
\sim \sigma \sim R_d$, we have $p(|R-R_\odot| \le d) \sim \frac{1}{2}
\textrm{erf}(\sqrt{2})\sim 0.48$.  Thus, the fraction of stars which
remain in the local volume in a diffusion time-scale strongly depends
on the ratio $d/\sigma$ and on the value of $R_d$.

The diffusion coefficient $D$ in Eqs.~(\ref{DiffEq})-(\ref{eqDiff})
can be regarded as an instantaneous coefficient which depends on the
position at which particles are initially localised and on the
diffusion time, since, as already mentioned, modeling the stellar
migration as a diffusion process is valid only for time intervals less
than the diffusion time-scale, $\Delta t \le T_D$.  As we described in
Brunetti et al.\ (\cite{brunetti}), the diffusion time-scale can be estimated
from the simulation results and it turns out to be of the same order
of the rotation period, $T_D \sim T_{\rm rot} = 2\pi/
\Omega(R,t)$. The diffusion coefficient is calculated by applying the
nonlinear least-square method which minimizes the difference between
the numerical results and the general solution of the diffusion
equation described by Eq.~(\ref{eqDiff}) for times less than $T_D$. 
At each time and radial position in our N-body simulations, we estimate the
instantaneous diffusion coefficient and related quantities, thus obtaining
a description of stellar migration along the whole simulation.
 
\section{Results}

\begin{figure*}
   \centering
   \includegraphics[width=7cm]{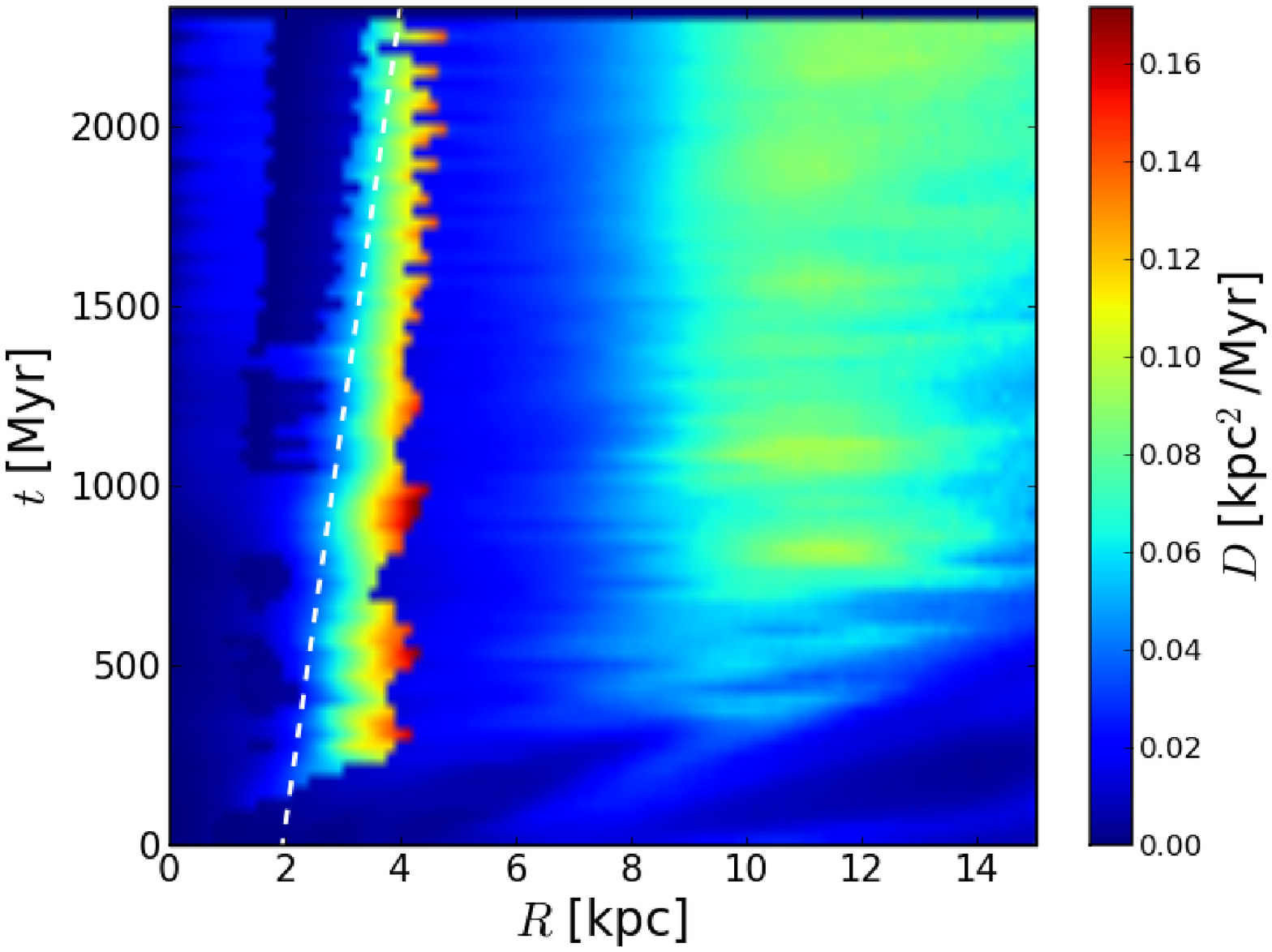}  
   \includegraphics[width=7cm]{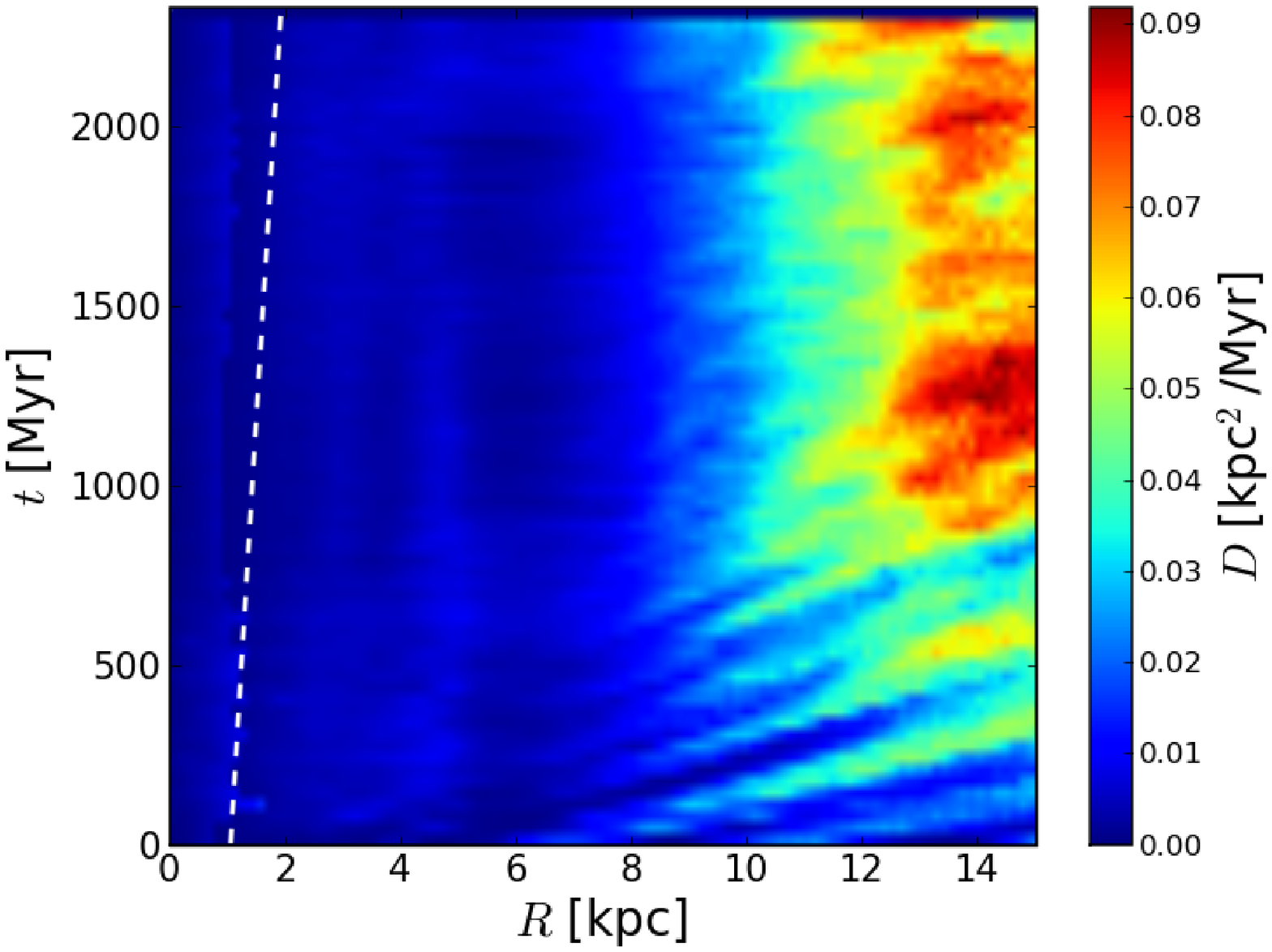} 
   \includegraphics[width=7cm]{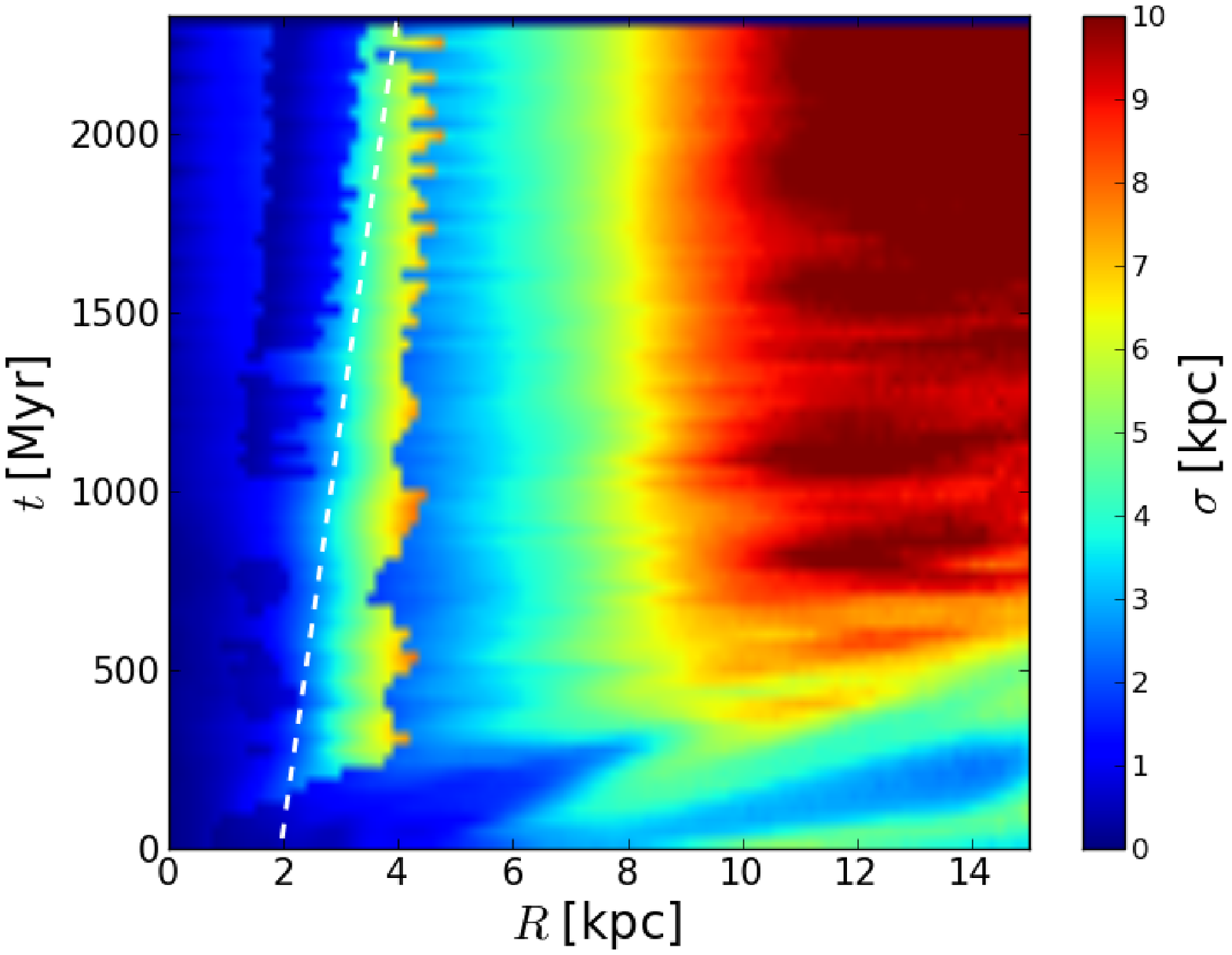}  
   \includegraphics[width=7cm]{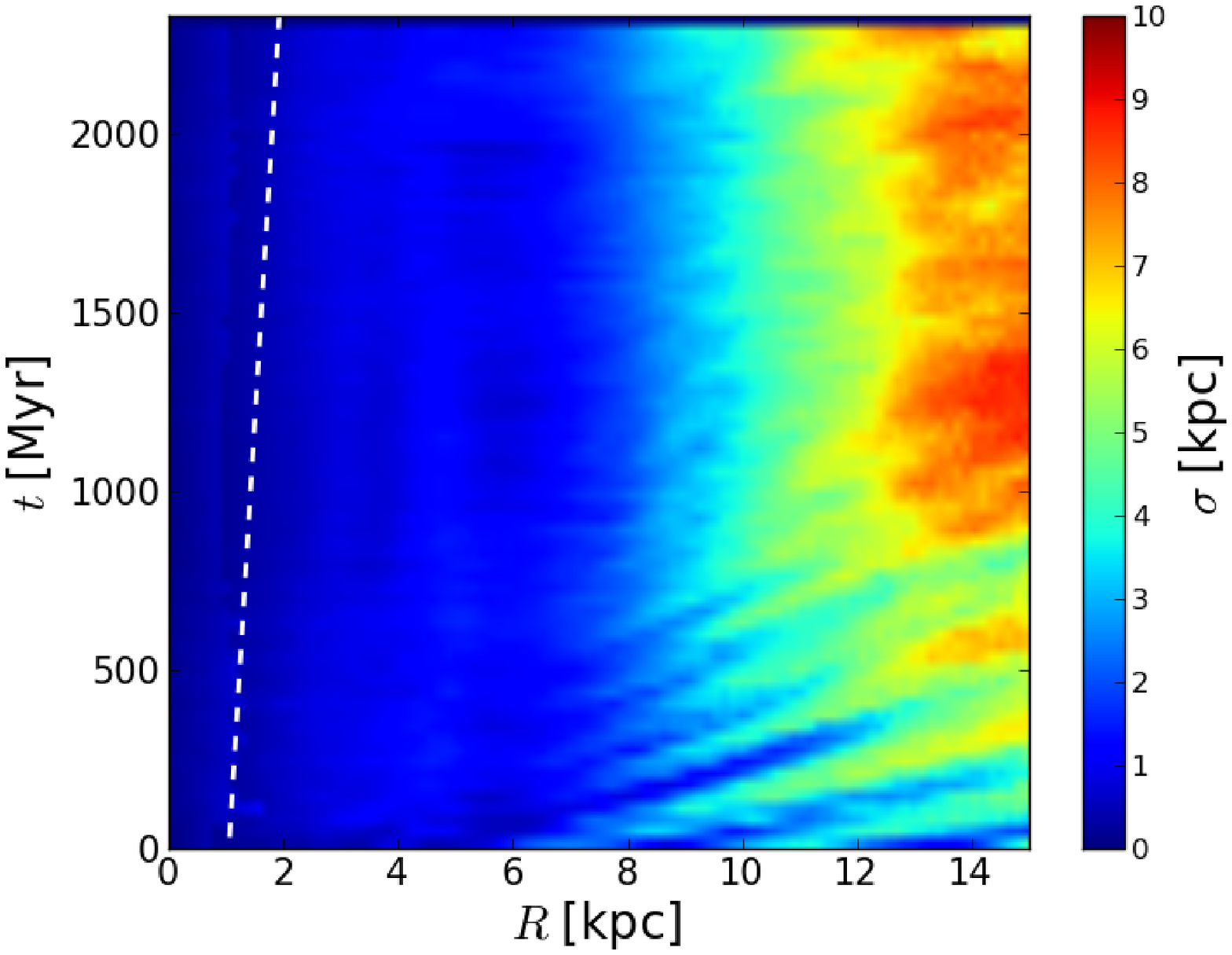}
   \includegraphics[width=7cm]{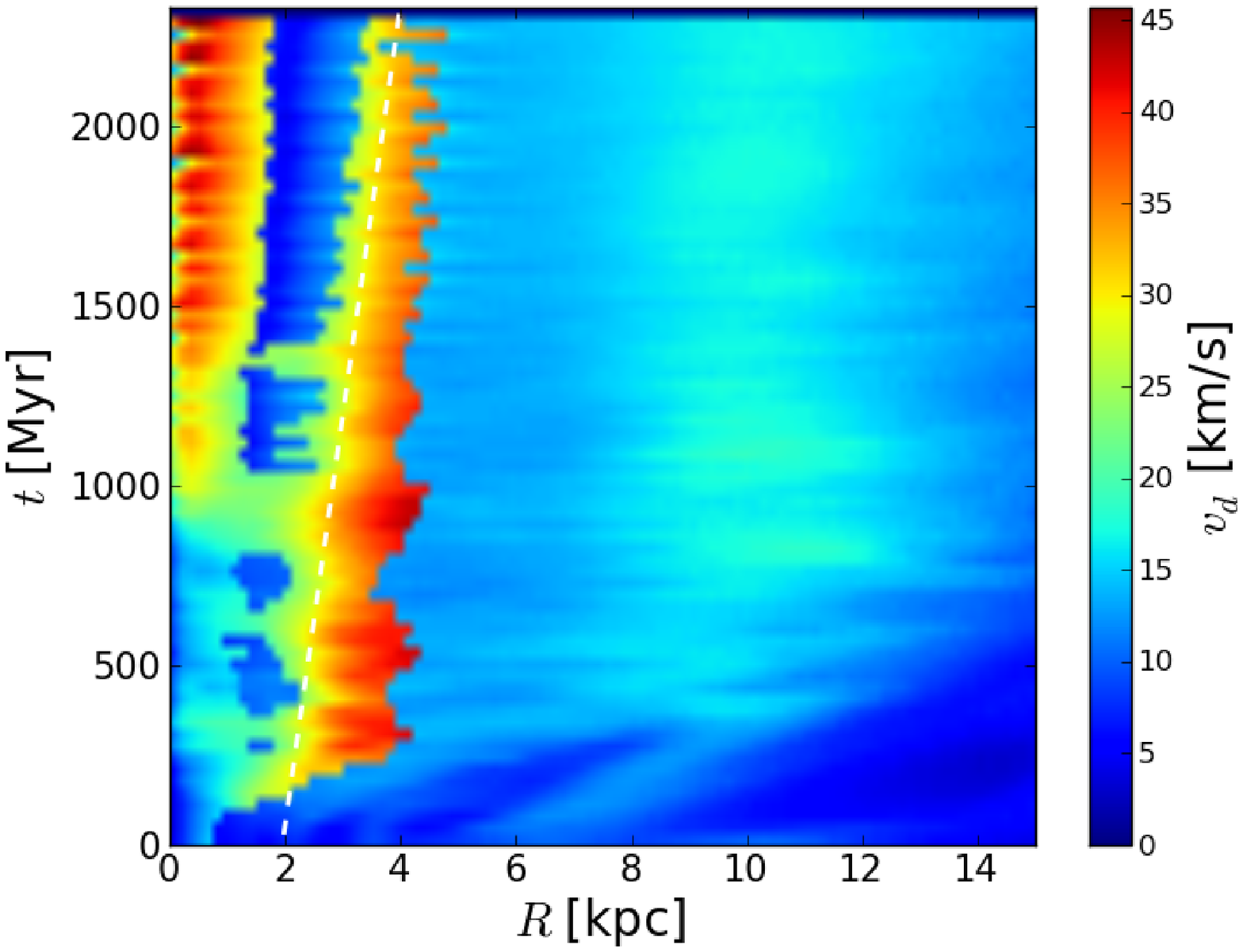}  
   \includegraphics[width=7cm]{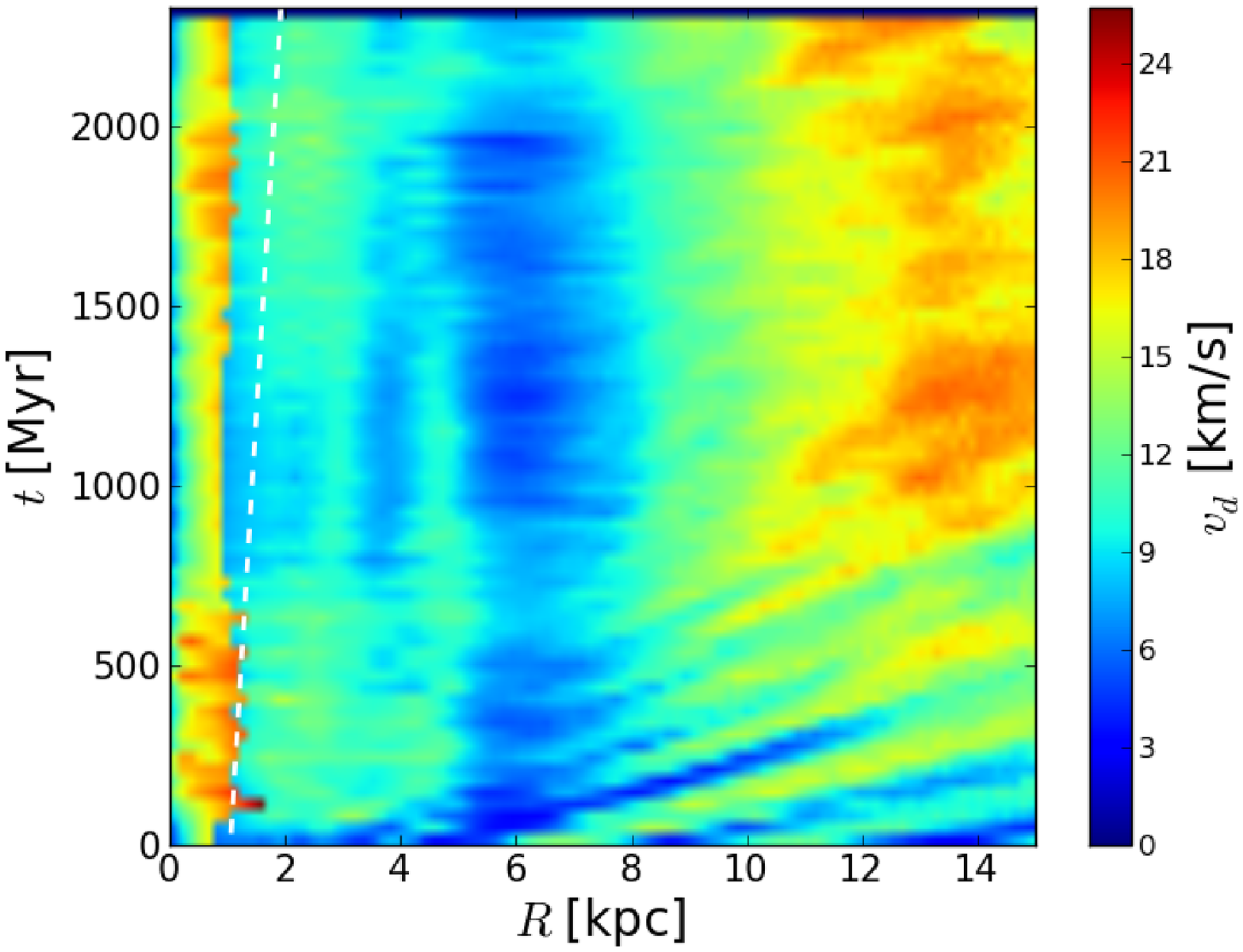}
   \caption{\small {\it Top row:} 
   Contour maps of the diffusion coefficient $D$. 
   {\it Middle row:} 
   Contour maps of the radial dispersion $\sigma = \sqrt{2\,D\, T_{rot}}$.
   {\it Bottom row:} 
   Contour maps of the diffusion velocity $v_D = \sigma/T_{rot} = \sqrt{2\,D/T_{rot}}$.
  {\it Left:} model {\tt m1}, {\it right:} model {\tt m6}.}   
    \label{Fig:kappa1}
\end{figure*}

The equations and the numerical methods described in the previous
section and in Brunetti et al.\ (\cite{brunetti}) allow us to calculate the
diffusion coefficient $D(R,t)$ which is shown in the contour maps of
Fig.~\ref{Fig:kappa1}, top row, for the models {\tt
  m1} (left panel) and {\tt m6} (right panel). The others models {\tt
  m2}, \ldots, {\tt m5} have intermediate values of the diffusion
coefficient between these two extreme cases.  The bar's corotation
radius is shown in Fig.~\ref{Fig:kappa1} as a dashed white line. It
can be seen that the diffusion coefficient is not constant in time nor
in radius. If the disk is not too hot, $D$ has the largest
values $D\sim 0.12~\textrm{kpc}^2\,\textrm{Myr}^{-1}$ 
outside the corotation radius of the bar (which
increases in time in our simulations), where the density is strongly
perturbed by a $m=2$ pattern created by the bar and the transient
spiral arms, and in the external regions $R>8$~kpc, where the density
is modulated by a $m=1$ pattern. The stars respond collectively to
these modulations and the process of migration corresponds to a
diffusion in an axisymmetric system. In hot disks, the diffusion
coefficient is large only in the external region $R>10$~kpc, where the
$m=1$ mode appears, with values of the order of $D\sim
0.08~\textrm{kpc}^2\, \textrm{Myr}^{-1}$. In this latter case, the
disk is not sufficiently cold to respond to the $m=2$ perturbation
created by the central bar.

\begin{figure*}
   \centering
   \includegraphics[width=7cm]{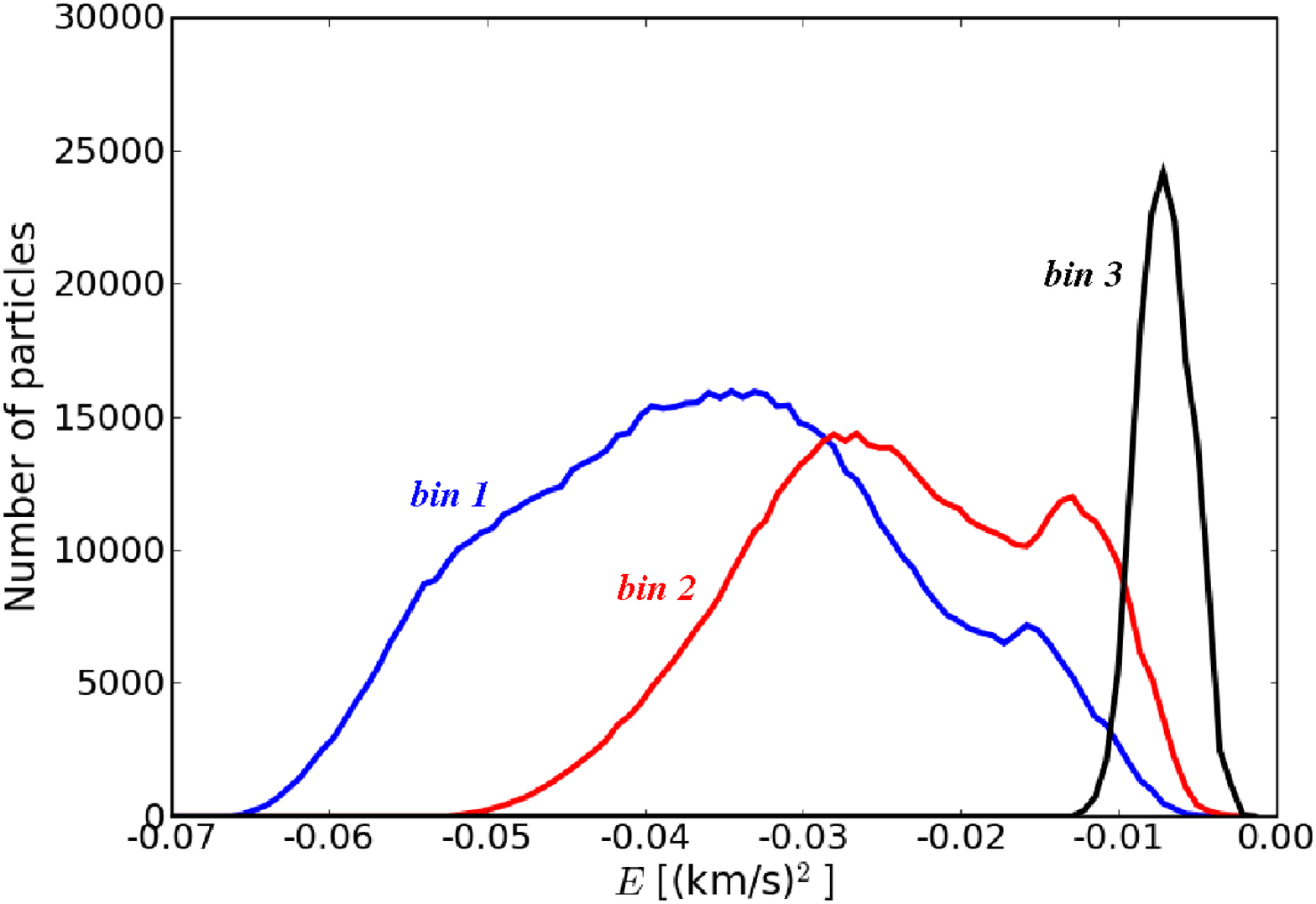}
   \includegraphics[width=7cm]{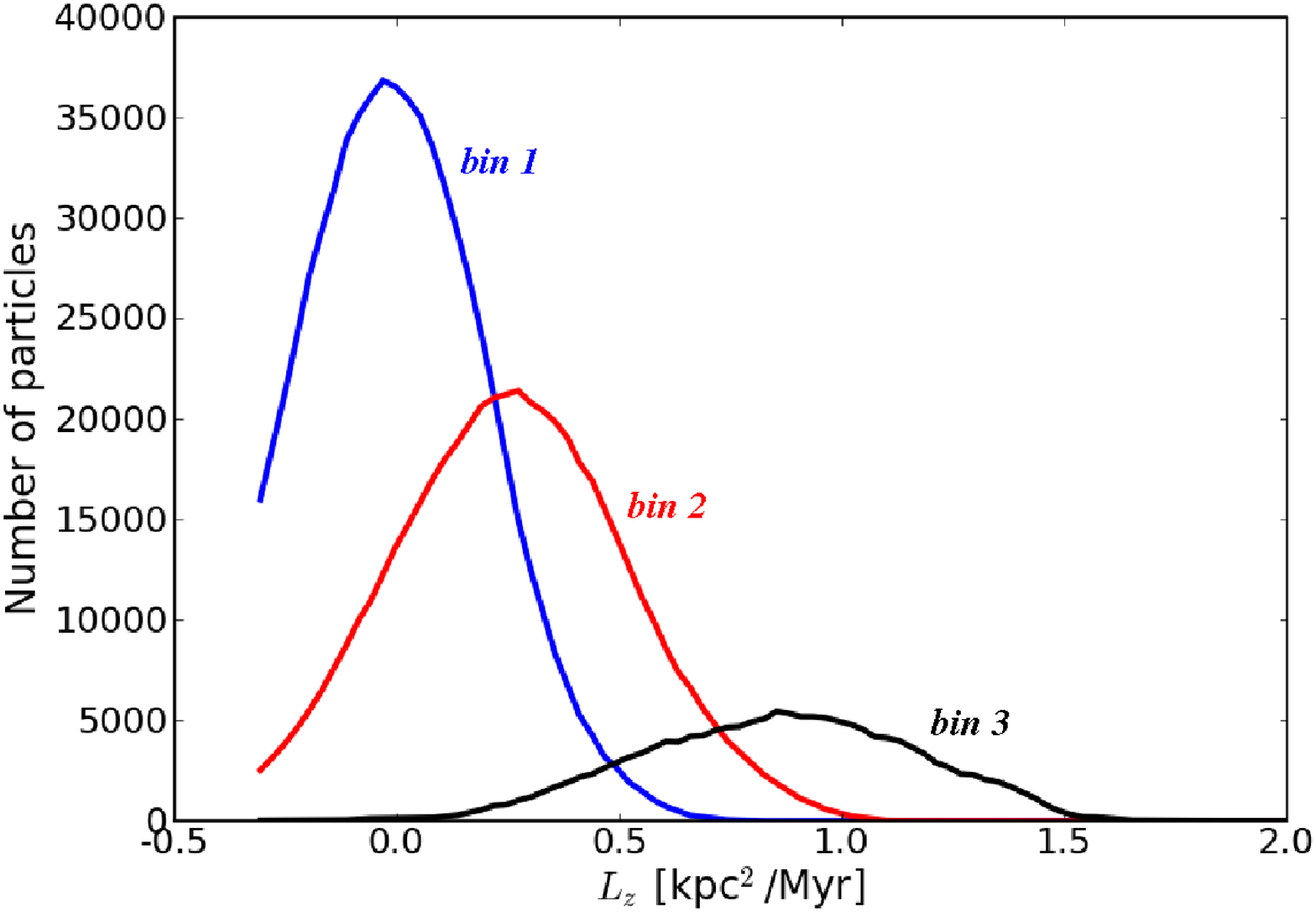}
   \caption{\small Energy values ({\it left panel}) and $L_z$-values 
({\it right panel}) of the particles at time 2.2~Gyr within the radial ranges 
$R=(1.5\pm 0.5)$~kpc (bin 1, blue lines), $R=(3.0\pm 0.5)$~kpc 
(bin 2, red lines) and $R=(8.0\pm 0.5)$~kpc (bin 3, black lines) for model 
{\tt m1}.} 
    \label{ELz}
\end{figure*}

When the disk is marginally stable with initial 
Safronov-Toomre parameter $Q_T \sim 1$, the diffusion coefficient
has the largest values just outside the corotation region, where two
different families of orbits are present, as can be inferred by the
total energy and angular momentum values of the particles in this
region.  We consider three different bins of particles located
respectively in $R=(1.5\pm 0.5)$~kpc, $R=(3.0\pm 0.5)$~kpc and
$R=(8.0\pm 0.5)$~kpc at $t=2.2$~Gyr in the model {\tt m1}. In the
first bin, particles are mainly inside the bar. Their energies $E$
span small negative values and the $z$-component of the angular
momentum is nearly centered in $L_z \sim 0$ (see the two panels of
Fig.~\ref{ELz}, blue lines, labeled as `bin 1'). Particles in the
second bin centered in $R=(3.0\pm 0.5)$~kpc at $t = 2.2$~Gyr belong to
two different types of orbits: one family can migrate only inside the
bar, the other can go outside the bar, in the disk. Large values of
the diffusion coefficient $D$ near the corotation region (see the left
panel of Fig.~\ref{Fig:kappa1}, top row) are related
to this superposition of two families of orbits.  The corresponding
values of $E$ and $L_z$ are shown in the panels of Fig.~\ref{ELz} (red
lines, labeled as `bin 2').  The two peaks in the energy distribution
correspond to the two families of orbits: bar particles have large
negative energies, while particles which can go into the disk have
small negative energies. Intermediate values correspond to the so
called hot particles, which as already mentioned before have a Jacobi
integral $H$ larger than the value of $H$ at the Lagrangian points
$L_{1,2}$, $H \ge H(L_{1,2})$.  We will discuss the hot particles
later on in this section.  It is important to note that the two peaks
are increasingly less evident in hotter disks and the number of hot
particles decreases as $Q_T$ increases (see the last column in
Table~1).  In the third bin, particles belong to the disk component:
these disk particles have small negative energies and large values of
$L_z$ (black lines, labeled as `bin 3').
 
From $D$ we can estimate the radial dispersion $\sigma$ in a rotation
period (which is of the order of the diffusion time-scale), $\sigma =
\sqrt{2\, D\, T_{\rm rot}}$. In the middle row of 
Fig.~\ref{Fig:kappa1} we show the contour maps of the radial 
dispersion for models {\tt m1} (left panel) and {\tt m6} (right 
panel). If the disk is sufficiently cold, the radial dispersion is
high near the corotation region of the bar, where it can recurrently
assume values of the order of $\sigma \sim 6$~kpc. This implies that
internal stars can recurrently be forced by the activity of the bar to
migrate in the external region of the disk. At an intermediate radius,
such as $R = 6$~kpc, the radial dispersion is of the order of $\sigma
\sim 3$~kpc and it increases in the external less dense regions where
the pattern $m=1$ dominates. In the external region, the diffusion of 
the stellar component is related to the presence of both patterns $m=1$ 
and $m=2$ and it can be enhanced in regions where the bar's outer Lindblad resonance 
overlaps with the spiral arms resonances (Minchev et al.~\cite{minchev11}).

The radial migration driven by the bar seems to be efficient in cold disks. 
According to our results for model {\tt m1}, the number of stars which stay
always in a local volume of 100~pc around $R=8$~kpc is low,
since $d \sim 100~\textrm{pc} \ll \sigma \sim 5$~kpc 
(see discussion after Eq.~(\ref{LocalVolume})). 
If the disk is hot, the values of $\sigma$ are lower than
the corresponding values for cold disks (at $R =
6$~kpc the radial dispersion is of the order of $\sigma \sim 1$~kpc 
for model {\tt m6} (see Fig.~\ref{Fig:kappa1}, middle row, right panel) 
and consequently radial migration is less effective than before.         

Another interesting quantity is the diffusion velocity in a rotation
period, defined by $v_{\rm D} = \sigma/T_{\rm rot} = 
\sqrt{2\, D/ T_{\rm rot}}$. We show the contour maps of this 
quantity in the bottom panels of Fig.~\ref{Fig:kappa1} for models 
{\tt m1} (left) and {\tt m6} (right). The diffusion velocity can
reach values of $v_{\rm D} \sim 40$~km/s near the corotation region in
the model {\tt m1}, while it is always less than $v_{\rm D}
\sim20$~km/s in the model {\tt m6}.

\begin{figure*}
   \centering
   \includegraphics[width=7cm]{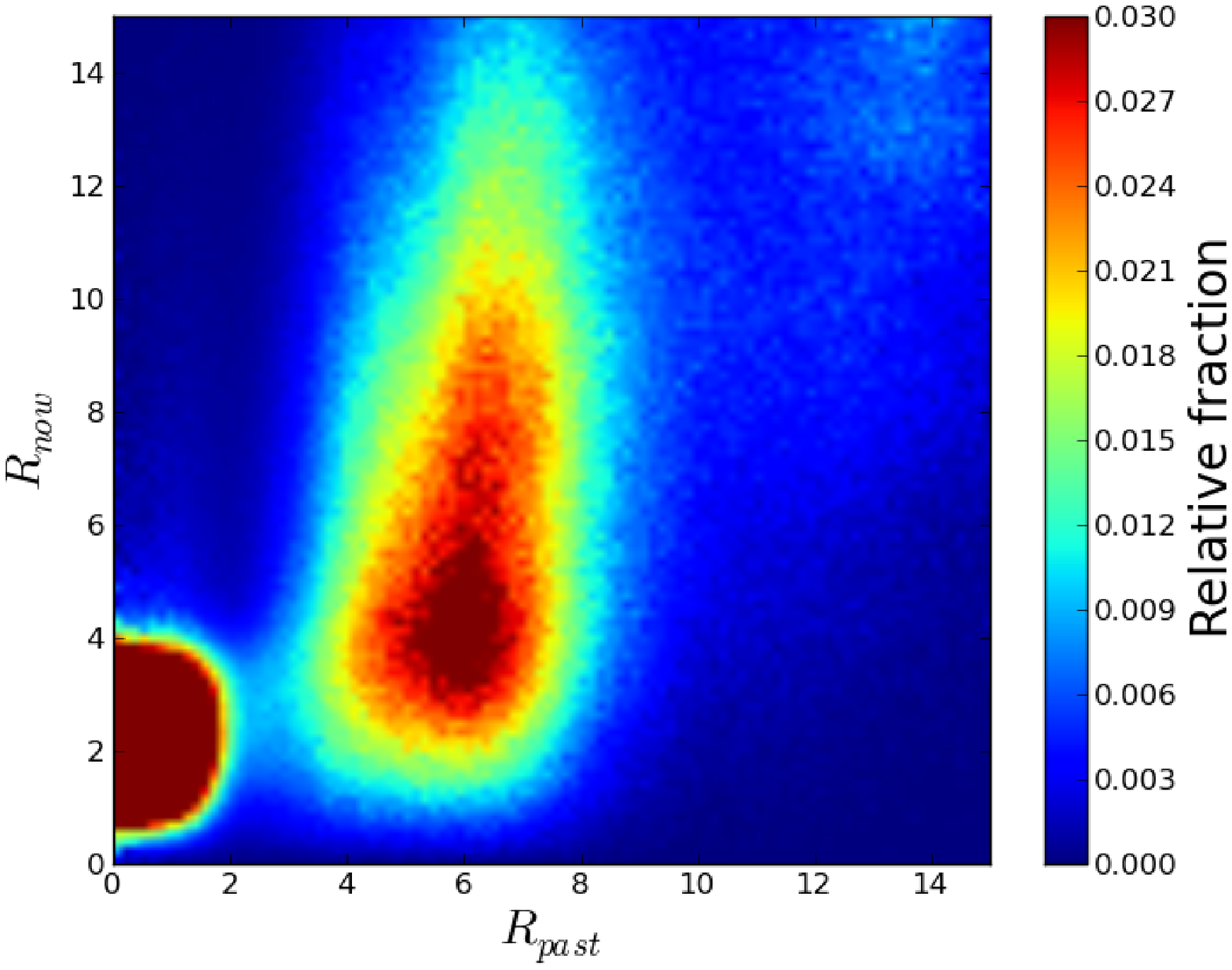}
   \includegraphics[width=7cm]{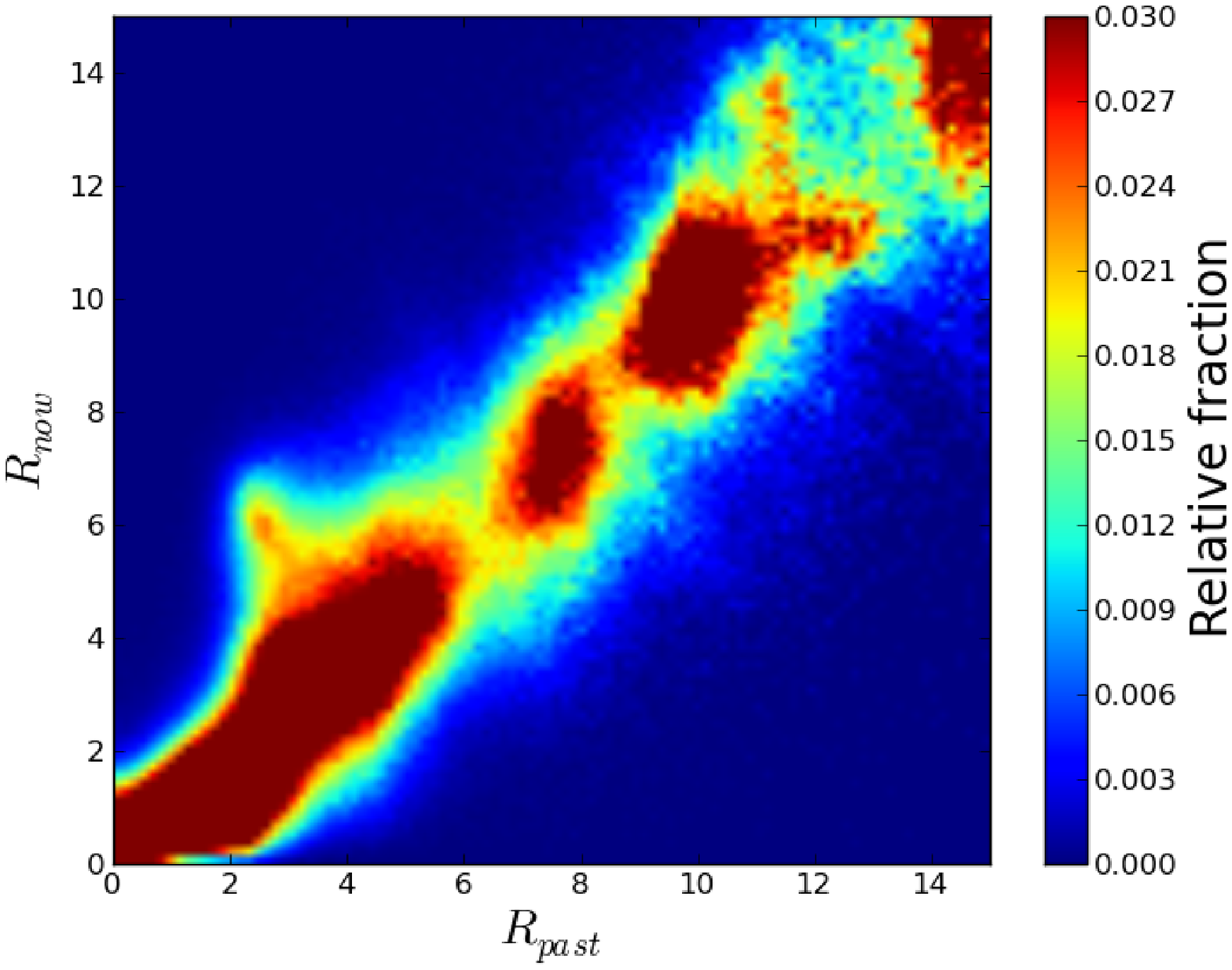}
   \caption{\small Relative fraction of stars which at time
     $t_{now}=2.2$~Gyr are at radial position $R_{\rm now}$ (vertical axis)
     and at time $t_{past} = 300$~Myr were at $R_{\rm past}$ (horizontal axis):
     ({\it left}) {\tt m1}, ({\it right}) {\tt m6}. }
    \label{Fig:R}
\end{figure*}

The fact that the corotation region plays a crucial role in stellar
diffusion can be also seen from Fig.~\ref{Fig:R}. Here it is shown how
the radial position of stars at some final time, 
$R_{\rm now}$ at $t_{now} = 2.2$~Gyr (on the vertical axis), depends on the
corresponding radial distribution $R_{\rm past}$ at time $t_{past} = 300$~Myr 
(on the horizontal axis). The different colors in the map
are related to the ratio of the number of stars at the two times,
$N(R_{past},t_{past})/N(R_{now},t_{now})$. It can be seen that in the case of model 
{\tt m1} (left panel) particles near the corotation radius, 
$R_{\rm now} \sim R_c = 4$~kpc (which is the value of the corotation 
radius at $t_{now} = 2.2$~Gyr in the considered model) 
came from inside and outside the corotation region, since they 
belong to two different families of orbits, as discussed before 
(see Figs.~\ref{ELz}). Particles outside the corotation at the final time 
were spread over the disk in the past, with $R_c <
R_{\rm past} \le 10$~kpc, while particles well inside the corotation radius, 
$R_{now} < 2$~kpc, were confined into the bar also in the past.
On the contrary, in the hot model {\tt m6} (see the
right panel of Fig.~\ref{Fig:R}) particles were 
essentially located at the same positions in the past,
with $R_{\rm past} \sim R_{\rm now} \pm \Delta R$, where $\Delta R \sim 1$~kpc 
and it slightly increases with $R_{\rm now}$. In this
case, no particular activity is observed in the corotation region,
$R_c \sim 2$~kpc.

\begin{figure*}
   \centering
   \includegraphics[width=6cm]{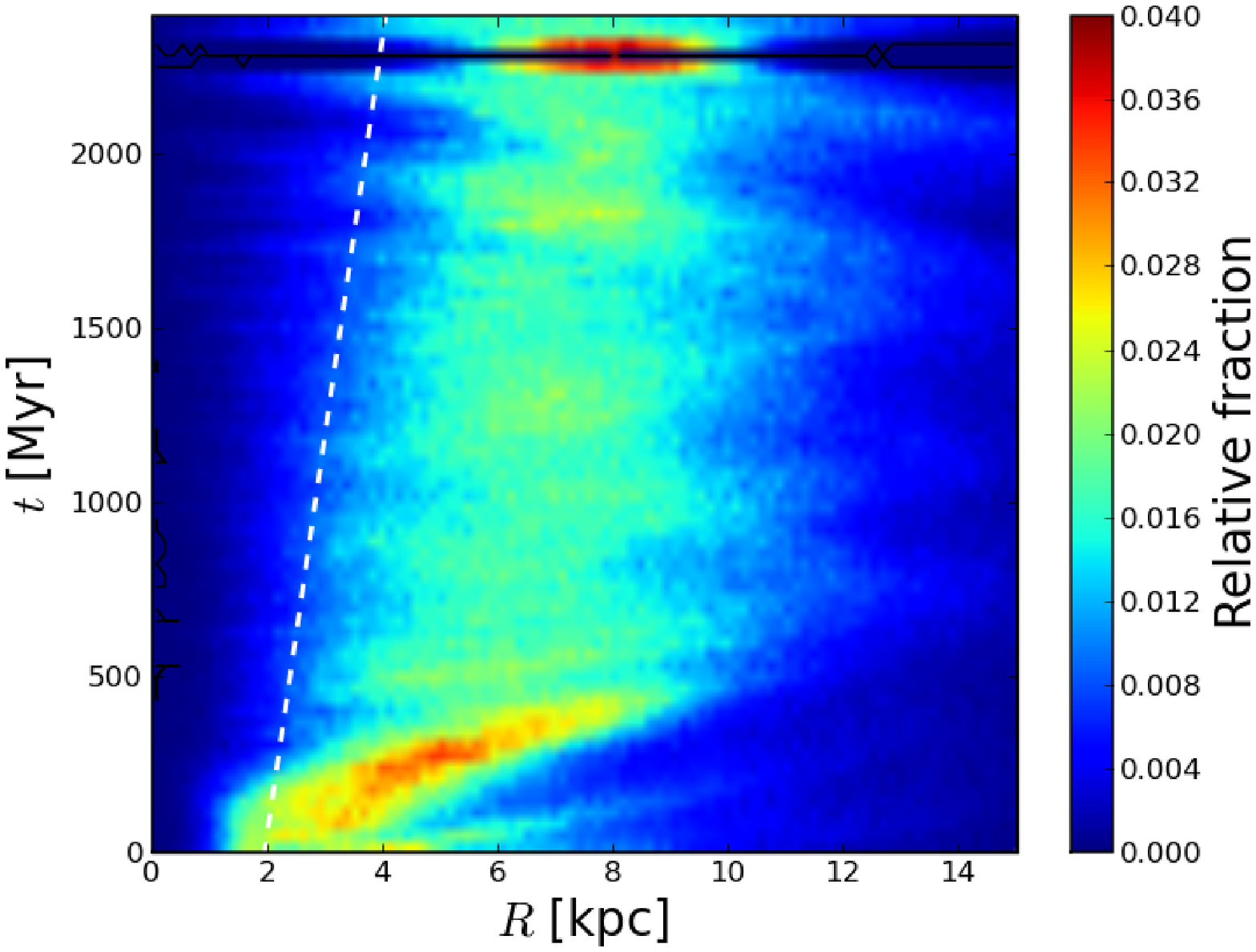}
   \includegraphics[width=6cm]{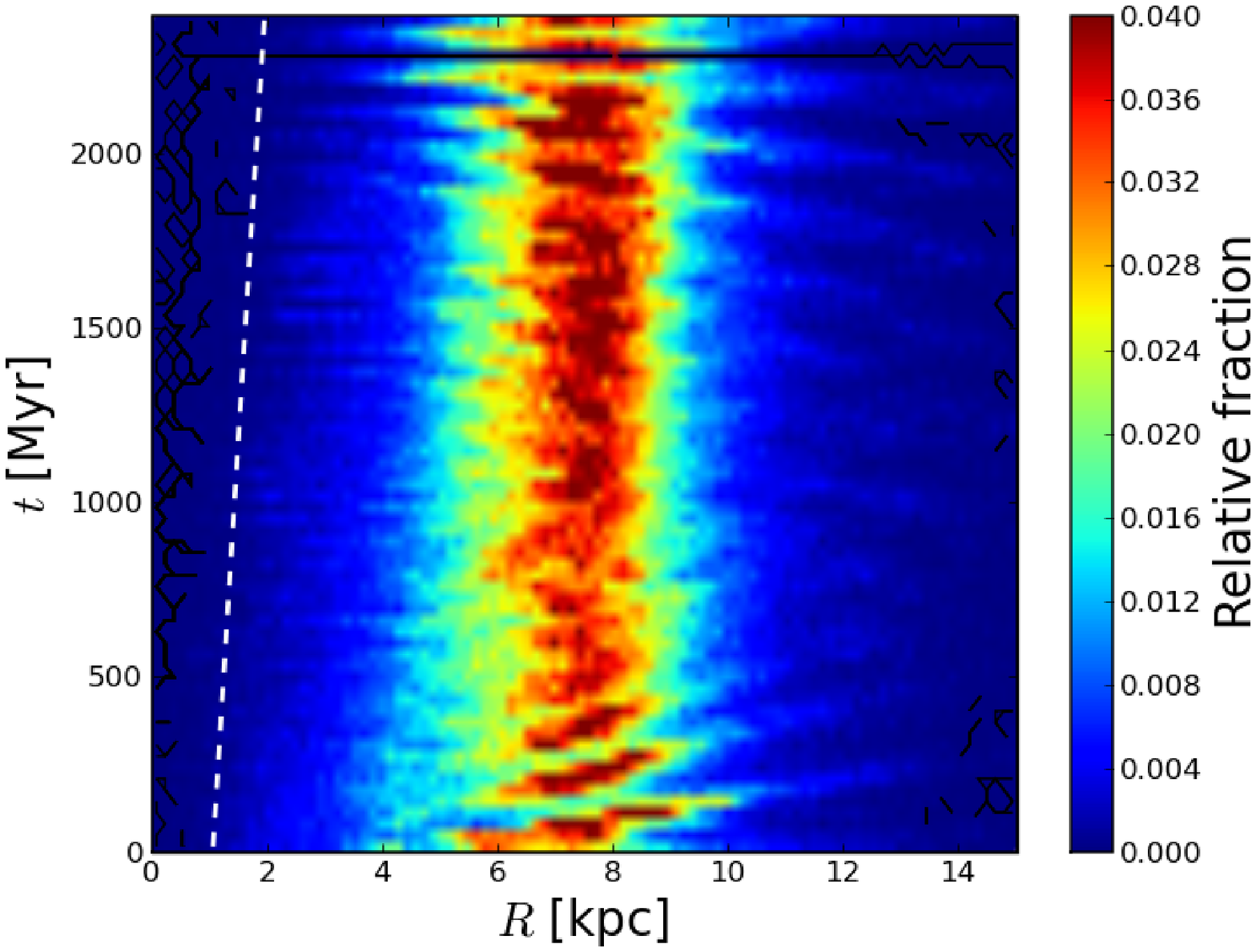}
   \includegraphics[width=6cm]{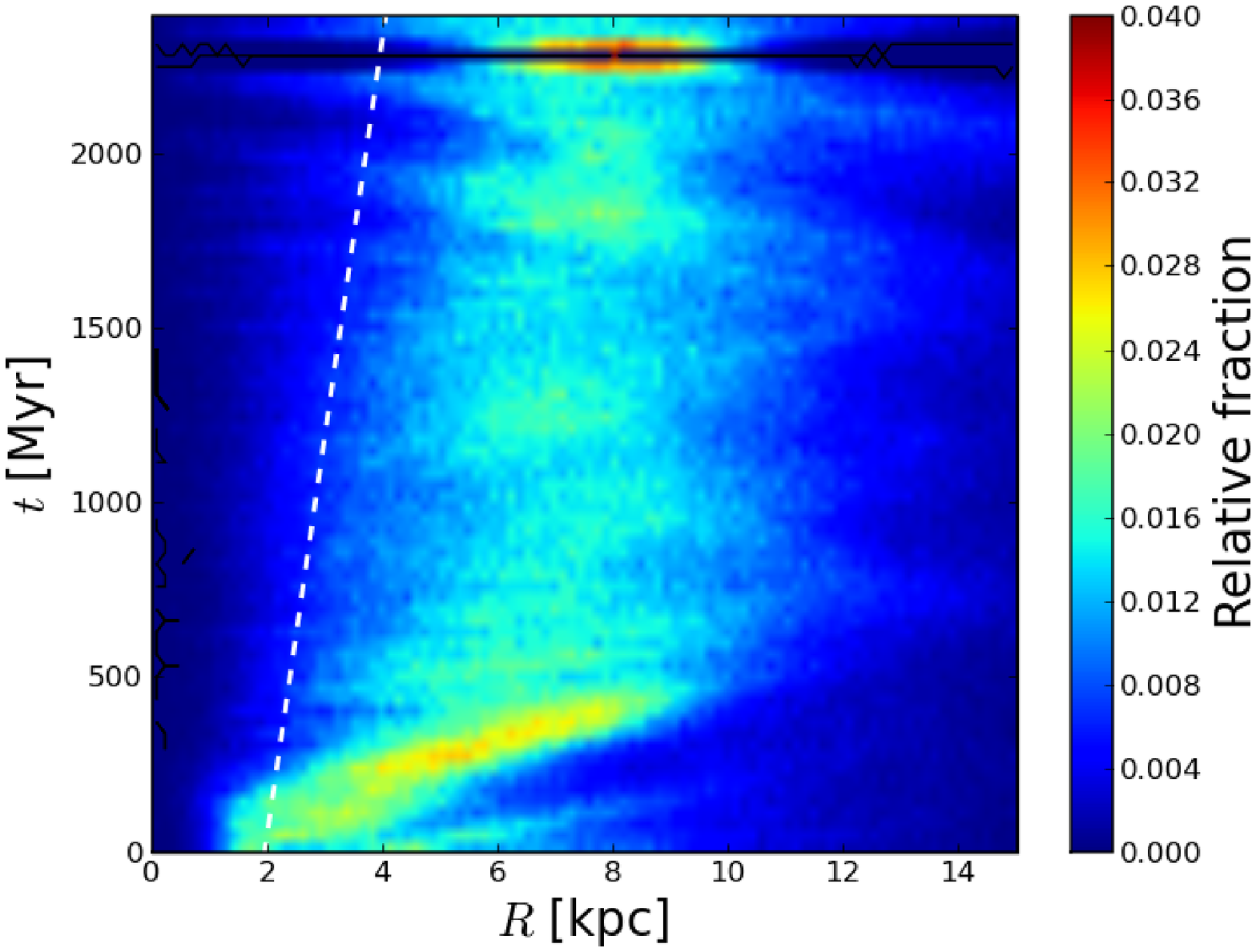}
   \caption{\small Evolution history of the radial position of stars 
   which at time $t_{now} =
   2.2$~Gyr are in $R_{now}=(8.0\pm 0.1)$~kpc: ({\it left}) {\tt m1}, 
   ({\it middle}) {\tt m6}, ({\it right}) {\tt m1} without hot particles.} 
   \label{Fig:RT}
\end{figure*}

As a further example, let us consider a bin of stars located
in $R_{now}=(8.0\pm 0.1)$~kpc at time $t_{now} = 2.2$~Gyr. Their
evolution history is shown in Fig.~\ref{Fig:RT}, for the two models,
{\tt m1} (left panel) and {\tt m6} (middle panel). When the bar's
strength is maximum (at $t \sim 350$~Myr, cf. Fig.~\ref{figC2Bar}),
stars localized near the corotation radius are forced to move toward
larger radii, since there the diffusion coefficient is large
(cf. Fig.~\ref{Fig:kappa1}, top row). Most of the
stars bounce then back and forth in the region $R_c < R<10$~kpc,
to reach the final bin position.  The evolution history in heated disks 
such as that of model {\tt m6} (see Fig.~\ref{Fig:RT},
middle panel) is very different, since stars are always localized in
the region $R = R_{now} \pm \sigma$, with $R_{now}=8$~kpc and
$\sigma \sim 2$~kpc, in agreement with the corresponding value of the
radial dispersion which can be inferred from the right panel of
Fig.~\ref{Fig:kappa1} (middle row).

Hot particles, characterized by a Jacobi integral $H>H(L_{1,2})$, have
a distribution which is maximum in the corotation region. In order to
investigate the role of such hot particles in the diffusion process,
we have removed them from the bin of stars localized in 
$R_{now}=(8.0\pm 0.1)$~kpc at time $t_{now} = 2.2$~Gyr which we
have considered just before. In the right panel of Fig. ~\ref{Fig:RT},
the evolution history of the bin where the hot particles have been
eliminated is shown. It must be compared with the evolution of the bin
which includes them, shown in the left panel of the same figure. It can
be seen that the main difference between the two panels is in the
relative number of stars which are able to reach the corotation
region. The hot particles migrate radially much more than the other
particles in the disk.  The number of hot particles in cold disks
(such as {\tt m1}) is nearly three times larger than the number in hot
disks (see last column in Table~\ref{table:1}).

\section{Discussion and conclusions}

From the previous sections it is clear that the amount of radial
migration in disk galaxies is strongly dependent on the bar and spiral
strengths. As we analyzed N-body simulations without gas, in order to
focus our study on the pure effect of the bar, we can only trace
radial migration for $\sim$2-3~Gyr. It is out of the scope of the
present study to compare our simulations with the observations of the
Solar neighborhood, which are the result of 10~Gyr of evolution.
Indeed, our models were not intended to reproduce the conditions in
our Galaxy, thus the precise values of, for example, the diffusion
time-scale $T_D$ and the radial dispersion $\sigma$ due to radial
migration obtained from these models do not correspond to those valid
for the Milky Way. However, the physical processes at play in cold
disks, like our model {\tt m1}, should be similar to those happening
in our Galaxy, where the Safronov-Toomre-parameter in the Solar
vicinity is $Q_T \sim 2$. Thus, we expect that the order of magnitude
of the relevant quantities obtained from our numerical results are
reasonably valid also for the earliest phases of the thin disk of our
Galaxy (within $\sim$ 2-3~Gyr from the bar formation).

\subsection{Implications for chemical evolution models of the Milky Way}

Chemical evolution models of our Galaxy traditionally assume that the
majority of stars do not migrate over large distances, and model the
Galaxy by introducing independent radial annuli which 
are wide enough (around 1-2~kpc wide) so that this approximation 
would be a valid one (van den Bergh~\cite{vandenbergh}; Schmidt~\cite{schmidt}; 
Pagel~\cite{pagel}; Chiappini et al.\ \cite{chiappini97}, Chiappini et al.\ 
\cite{chiappini01}). The expectation is
that {\it intruders} from other galactocentric distances would not
represent more than a few percent of the stars in the local
samples. However, as discussed in Sect.~1, there are recent claims
that radial migration was more efficient than previously
assumed. This, in turn, is driven by the large scatter in the AMR of
the Geneva-Copenhagen sample.

In the particular model of Chiappini et al. (\cite{chiappini01}), each annulus 
is 2~kpc wide (i.e. $d=2$~kpc). Thus, in this case in the Solar
vicinity $d$ is nearly half the value of the radial dispersion
obtained in model {\tt m1} at intermediate radii 6-8~kpc,
$d\sim 2\sigma$.  
Thus, the relation between the different 
length-scales which is approximately valid in the Solar vicinity 
is $d \sim \sigma \sim R_d$ (see discussion after 
Eq.~(\ref{LocalVolume})) and the percentage of stars which stayed in 
a volume $|R-R_\odot| \le d = 2$~kpc turns out to be of the order of 
50\% in a diffusion time-scale, $T_D \sim T_{\rm rot} = 2\pi 
R_\odot/V_{\rm c} \sim 223$~Myr near the Sun. The region from which
the rest of stars comes from depends on the activity of the bar which
is not a constant pattern, as assumed in the past (Wielen~\cite{wielen}).
If we consider for example the recurrent bar scenario described in
Bournaud \& Combes (\cite{bournaud}), the bar can be rebuilt several times in a
Hubble time in galaxies with significant gas accretion.  When the bar
strength is high, such as at the beginning of our simulations and in
general when the bar is rebuilt by episodes of dissipative infall of
gas, stars come mainly from the corotation region (see
Fig.~\ref{Fig:RT}, left panel, $t<500$~Myr), which is closer to the
center since the corotation radius typically becomes smaller after
each reformation episode (Bournaud \& Combes~\cite{bournaud}). When the bar
strength saturates to a constant value during a quiescent phase, stars
can span the region between the corotation radius and $\sim$10-11~kpc 
from the galactic center (see Fig.~\ref{Fig:RT}, left panel,
$t>500$~Myr).

\begin{figure*}
   \centering
   \includegraphics[width=7cm]{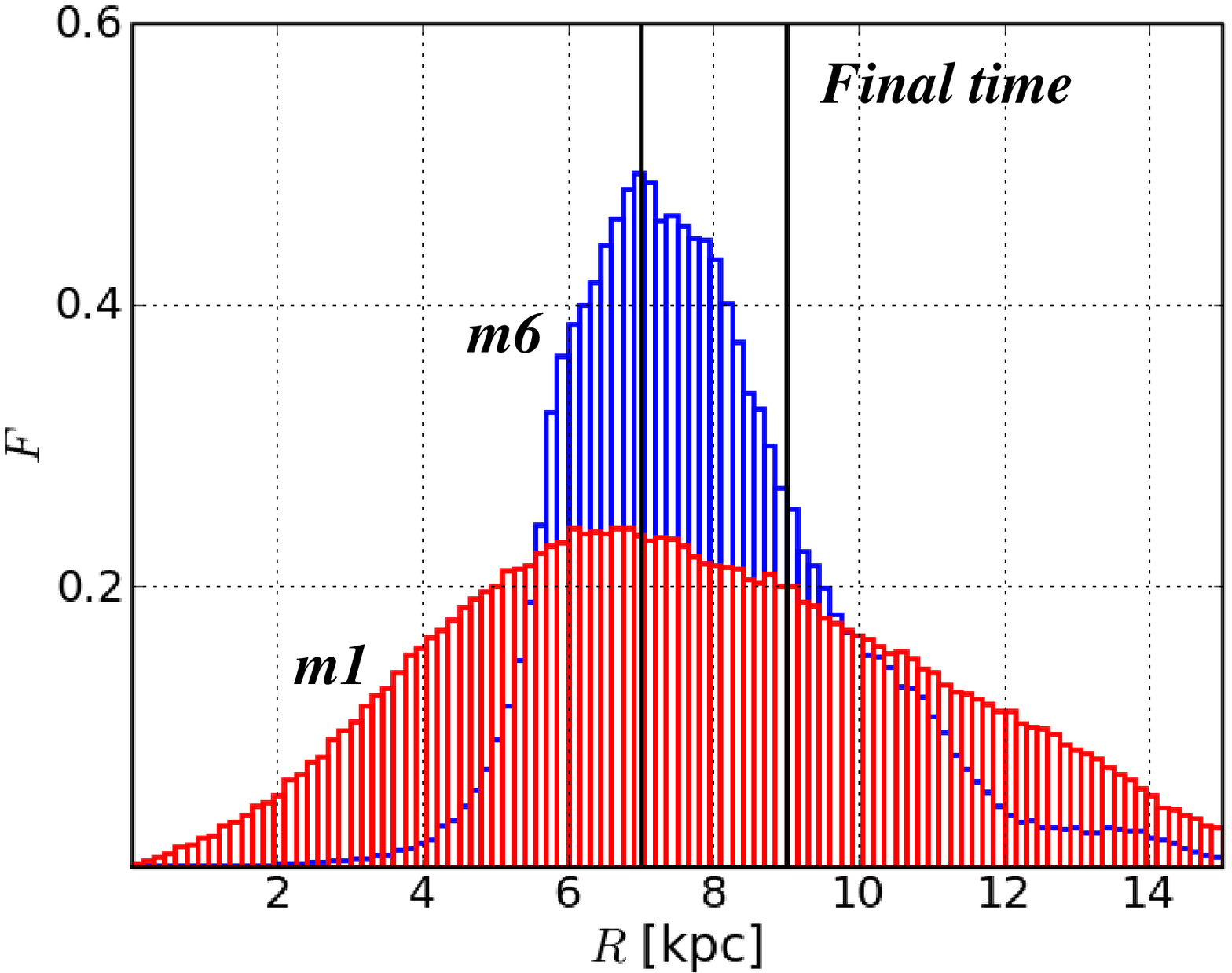}
   \includegraphics[width=7cm]{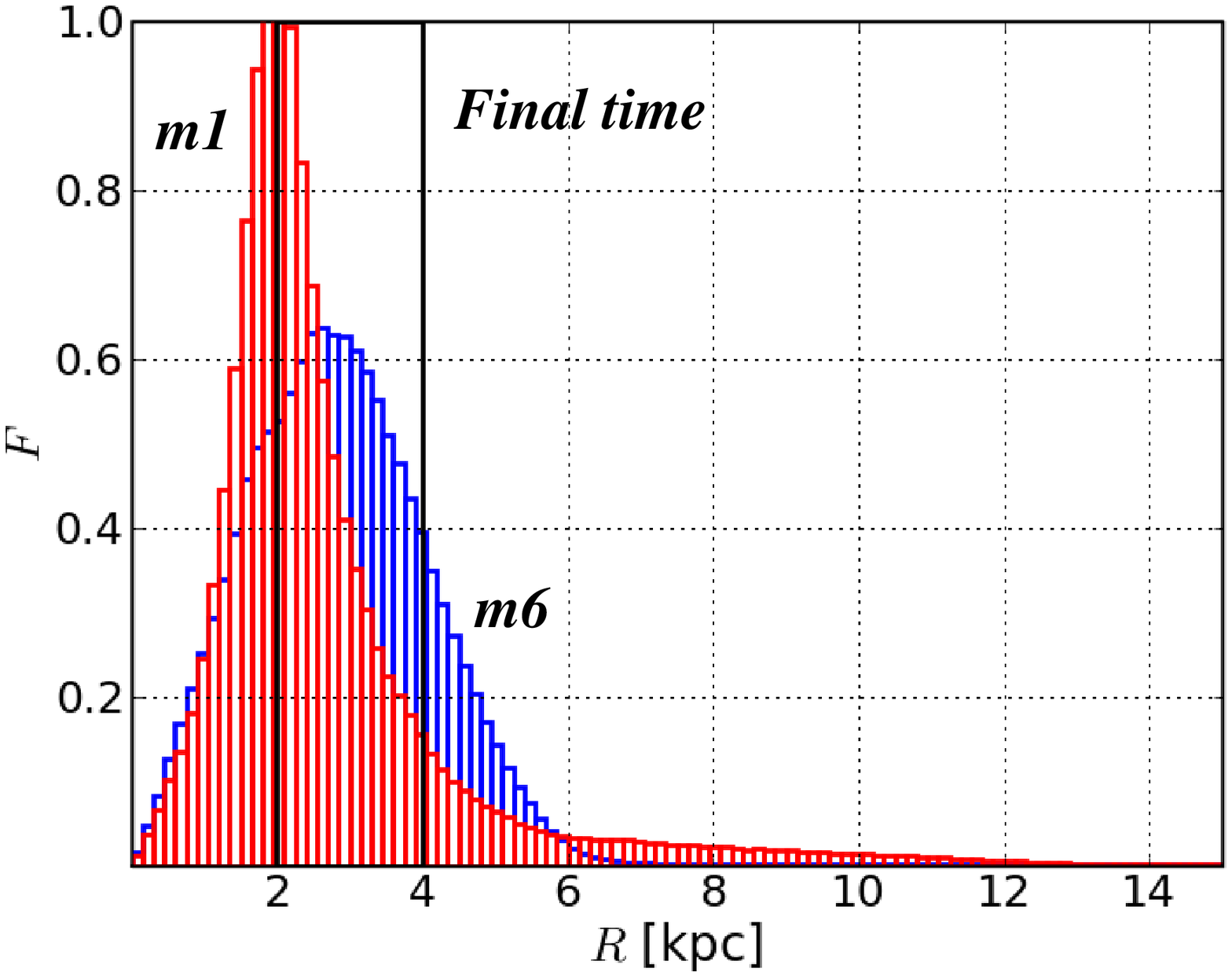}
   \caption{\small  Histograms of radial position of stars in the bins 
$R= (8\pm 1)$~kpc ({\it left panel}) and $R= (3\pm 1)$~kpc ({\it right panel}) 
at the final time (black line) and $2.2$~Gyr before (red for model {\tt m1} 
and blue for model {\tt m6}).}
      \label{Fig:Examples}
\end{figure*}

In Fig.~\ref{Fig:Examples} we show two examples of stellar
diffusion. Stars localized in $R= (8\pm 1)$~kpc and $R= (3\pm 1)$~kpc
at the end of the simulations are in the black bins, in the left and
right panel, respectively. The red and blue distributions correspond
to the radial positions of the stars $2.2$~Gyr before, for model {\tt
  m1} and {\tt m6}, respectively. As can be seen from the left panel
in Fig.~\ref{Fig:Examples}, when the disk is sufficiently cold (model
{\tt m1}, red distribution), the radial dispersion is higher than the
corresponding values for the hot disk (model {\tt m6}, blue
distribution). In 2.2~Gyr (which is much larger than the diffusion
time-scale), only 25\% of the stars remain in the black bin, all the
others come from outside.  When the disk is hot, the percentage is
nearly twice. In the right panel we show the case of stars near the
corotation region (remember that the corotation radius is $R_c= 2$~kpc
for model {\tt m1} and 1~kpc for model {\tt m6}). Stars come mainly
from the corotation radius when the disk is cold (red distribution),
while more than 60\% stay at the same position if the disk is hot.

We can conclude that the dynamical effects of stellar migration should
be included in chemical evolution models of our Galaxy insofar as the
distance $d$ between each annulus is smaller than the radial
dispersion $\sigma$, which is related to the diffusion coefficient $D$
and to the diffusion time-scale $T_D$ by $\sigma = \sqrt{2\, D\,
  T_D}$. The radial dispersion $\sigma$ depends on the degree of
marginality of the disk.

\subsection{Conclusions}

Disk galaxies are complex systems where collective phenomena give rise
to the emergence of nonlinear structures, such as central bars and
spiral arms.  We have investigated the role of bars in marginally
stable disks and overheated disks, and their forcing effect on the hot
(chaotic) particle component, which is more sensitive to
external/internal perturbations.

Modeling the migration of stars in marginally stable disks as a
diffusion process in the radial direction is a powerful tool which
allows us to estimate quantitatively two crucial parameters, the
diffusion coefficient and the diffusion time-scale. With these 
quantities we are able to compute two other fundamental quantities 
which are the radial dispersion and the diffusion velocities at 
different radii and at different times. It is important to note
that such a diffusion model makes sense only if there is a
stochastic microscopic component at the origin of the
diffusion. Ideally, the diffusion time-scale should 
not be much shorter than the microscopic e-folding time due to 
chaos, which in $N$-body systems is typically of the order of the 
dynamical time (Miller~\cite{miller}). Over longer time-scales, the diffusion 
model becomes influenced by the mere global dynamical evolution of 
the disk, so its behavior may depart from a simple linear diffusion 
equation with constant coefficient. Thus, the present diffusion 
calculation is useful for diffusion time-scales in a range around 
the rotational period. At each time and radial 
position in $N$-body simulations, we are able to estimate the 
instantaneous diffusion coefficient and the related quantities 
(i.e., diffusion time-scale, radial dispersion and diffusion 
velocity), thus obtaining a description of the stellar diffusion on 
the whole simulation.
 	
We have found that the diffusion time-scale is of the order of one
rotation period and that the diffusion coefficient $D$ depends on the
evolution history of the disk and on the radial position. Larger
values $D$ are found in cold disks near the corotation region, which evolves in
time, and in the external region, where asymmetric patterns develop.
Marginally stable disks, with $Q_T \sim 1$, have two different
families of bar orbits with different values of angular momentum $L_z$
and energy $E$, which determine a large diffusion in the corotation
region.  In hot disks, $Q_T > 1$, stellar diffusion is much
more reduced than in the case of marginal disks.

The calculations of both the diffusion coefficient and the diffusion
time-scale give us a quantitative measure of the migration process in
the disk. Another advantage of studying the diffusion of stars in real
space, rather than in velocity space, is that it can be more easily related
to the evolution of chemical elements, which can be modeled as tracers
which follow the evolution of the stellar component.  The diffusion
process of stars and tracers can be directly implemented in chemical
evolution codes, which we plan to do in the near future.

It is interesting to compare our results with those obtained in a recent analysis of Shevchenko (\cite{shevchenko}), where the Lyapunov and the diffusion times are estimated for the Quillen's model (Quillen~\cite{quillen2003}) which describes the Hamiltonian motion in the Solar neighborhood due to the interactions of bar and spiral arms resonances. He found that the Lyapunov time, of the order of 10~Galactic years, depends weakly on the model parameters, which can radically change the extent of the chaotic domain (as we obtain by varying the Safronov-Toomre parameter $Q_T$). The diffusion time, which characterizes the transport in the chaotic domain of the phase space, is calculated as the inverse of the diffusion rate in the energy variable (thus differing from our definition), with upper bounds of the order of 10~Gyr. It strongly depends on the radial position in the Galaxy, in agreement with our findings.

We call attention to the fact that although there are good reasons 
(both theoretical and observational) to expect that radial process 
took place during the evolution of our Galaxy, it could have been 
much weaker than what has been proposed so far, as implied by the 
existence of radial abundance gradients, and its mild time-variation. 
Unfortunately, the uncertainties in the observed abundance gradient 
evolution are still large and we need to wait until better ages and 
distances will be available, which will happen in the near future 
thanks to GAIA and asteroseismology.

Meanwhile, the theoretical work should focus on the relative 
importance of the main drivers of radial migration, and in the case of 
the Milky Way, on the role of the bar. Here we have shown that the 
radial migration process is not only time-dependent but also changes 
with galactocentric distance, in connection to the bar, plus spiral arms.
We plan to analyze simulations with bar but including also gas accretion, 
where the radial migration process can be traced for several Gyrs. 
In this way we will be able to answer if radial migration could have 
had repeated peaks along the MW history or if it faded away after the 
first 2-3~Gyr. These models coupled with the chemical information 
will be essential to interpret the radial mixing effects on the local 
age-metallicity relation, the metallicity distributions of the local 
thick and thin disks, and on the evolution of the abundance gradients 
in both disks.

\begin{acknowledgements}
  Simulations have been run on the REGOR cluster at Geneva
  Observatory. We thank Michel Grenon and Ralph Sch\"onrich for useful
  discussions.  This work has been supported by the Swiss National
  Science Foundation.
\end{acknowledgements}

\end{document}